\documentclass{aa}
\usepackage[dvipsnames,table]{xcolor}
\usepackage{graphicx,colortbl}
\usepackage{natbib}
\newcounter{rowno}
\usepackage{tabularx}
\usepackage{txfonts}
\usepackage{hyperref}
\hypersetup{
    colorlinks=true,
    linkcolor=blue,
    citecolor=blue,
    urlcolor=blue
}
\graphicspath{{figures/}}
\DeclareMathAlphabet{\mathitbf}{OML}{cmm}{b}{it}
\newcommand{\vA}{\mathitbf{A}}
\newcommand{\vAp}{\mathitbf{A}_{\mathrm{0}}}
\newcommand{\vB}{\mathitbf{B}}
\newcommand{\vJ}{\mathitbf{J}}

\newcommand{\vBj}{\mathitbf{B}_{\rm J}}
\newcommand{\vBh}{\mathitbf{B}_{\rm h}}
\newcommand{\vBp}{\mathitbf{B}_{\mathrm{0}}}
\newcommand{\vBpn}{\mathitbf{B}_{\mathrm{0},n}}

\newcommand{\dV}{\mathrm{d}\mathit{V}}

\newcommand{\Ef}{E_{\rm F}}

\newcommand{\Efpre}{E_{\rm F,pre}}
\newcommand{\Efpost}{E_{\rm F,post}}
\newcommand{\Ep}{E_0}
\newcommand{\Et}{E}

\newcommand{\Efnt}{\Ef/\Et}
\newcommand{\Efnp}{\Ef/\Ep}

\newcommand{\Ediv}{E_{\rm div}}
\newcommand{\Edivprime}{E_{\rm div}/E}
\newcommand{\Fabs}{\phi_m}

\newcommand{\thetaj}{\theta_J}
\newcommand{\Hj}{H_{\rm J}}
\newcommand{\Hpj}{H_{\rm PJ}}
\newcommand{\Hv}{H_{\rm V}}
\newcommand{\Hvpre}{H_{\rm V,pre}}
\newcommand{\Hvpost}{H_{\rm V,post}}

\newcommand{\Hvabs}{|H_{\rm V}|}

\newcommand{\Hjabs}{|H_{\rm J}|}

\newcommand{\Hjn}{|\Hj|/|\Hv|}

\newcommand{\Hjfn}{|\Hj|/\tilde\phi^2}

\newcommand{\Hvfn}{|\Hv|/\tilde\phi^2}

\newcommand{\alphat}{\alpha_{\rm tot}}
\newcommand{\Rprime}{R/\tilde\phi}

\newcommand{\hcrit}{h_{\rm crit}}
\newcommand{\dfc}{d_{\rm  FC}}
\newcommand{\dpc}{d_{\rm PC}}
\newcommand{\dfcprime}{\dfc/(\dpc/2)}
\newcommand{\Csr}{C_{\rm SR}}
\newcommand{\Ccum}{{C_{\rm cum}}}
\newcommand{\mCcum}{\langle C_{\rm cum}\rangle}
\newcommand{\curlA}{\nabla\times\vA}
\newcommand{\curlAp}{\nabla\times\vAp}
\newcommand{\divB}{\nabla\cdot\vB}
\newcommand{\divBp}{\nabla\cdot\vBp}

\newcommand{\cerg}{C_{\rm erg}}
\newcommand{\mxmx}{{\rm Mx}^2}
\newcommand{\mx}{{\rm Mx}}

\newcommand{\erg}{{\rm erg}}
\newcommand{\cme}{E}
\newcommand{\Sall}{S_{\rm M1+}}
\newcommand{\Sminor}{S_{\rm M4-}}
\newcommand{\Smajor}{S_{\rm major}}
\newcommand{\Ssea}{S_{\rm SEA}}
\newcommand{\Scbetal}{S_{\rm CBetal}}
\newcommand{\Sxonly}{S_{\rm X1+}}
\newcommand{\ie}{{i.e.}}
\newcommand{\eg}{{e.g.}}

\newcommand{\gray}{\color{gray}}
\newcommand{\scol}{\cellcolor{red!50}}
\newcommand{\mcol}{\cellcolor{orange!50}}
\newcommand{\wcol}{\cellcolor{orange!25}}
\begin{document}

\title{Conditioning of the solar corona due to large flares}
\author{J.K. Thalmann\inst{1}
        \and M. Gupta\inst{1}
 \and A.M. Veronig\inst{1,2}
 \and Y. Liu\inst{3}
 }
\institute{University of Graz, Institute of Physics, Universit\"atsplatz 5, 8010 Graz, Austria\\
 \email{julia.thalmann@uni-graz.at}
 \and University of Graz, Kanzelh\"ohe Observatory for solar and Environmental Research, Kanzelh\"ohe 19, 9521 Treffen, Austria
 \and W.W. Hansen Experimental Physics Laboratory, Stanford University, Stanford, CA 94305-4085, USA
 }

\date{Language revised \today; accepted January 2, 2025}
\titlerunning{Conditioning of the solar corona due to large flares}

\abstract
{}
{
We aim to better characterize the conditions of the solar corona, especially with respect to the occurrence of confined and eruptive flares. In this work, we model the coronal evolution around 231 large flares observed during solar cycle~24.
}
{
Using Helioseismic and Magnetic Imager vector magnetic field data around each event, we employed nonlinear force-free field extrapolations to approximate the coronal energy and helicity budgets of the solar source regions. A superposed epoch analysis and dynamical time warping applied to the time series of selected photospheric and coronal quantities were used to pin down the characteristics of the pre- and postflare time evolution, as well as to assess flare-related changes.
}
{
During the 24~hours leading up to a major flare, the total magnetic energy and unsigned magnetic flux were seen to evolve closely with respect to each other, irrespective of the flare type. Prior to confined flares, the free energy evolves in a way that exhibits more of a similarity with the unsigned flux than the helicity of the current-carrying field, while the opposite trend is seen prior to eruptive flares. Furthermore, the flare type can be predicted correctly in more than 90\% of major flares when combining measures of the active regions nonpotentiality and local stability.  The coronal energy and helicity budgets return to preflare levels within $\approx$\,6 to 12~hours after eruptive major M-class flares, while the impact of eruptive X-flares lasts considerably longer. Finally, the postflare replenishment times of $\gtrsim$12~hours after eruptive X-class flares may serve as a partial explanation for the rare observation of eruptive X-class flares within a time frame of a few hours.
}
{}
\maketitle

\section{Introduction}
\label{s:intro}

Solar flares and coronal mass ejections (CMEs) are often caused by the interaction of magnetic field in coronal loops rooted in regions of strong magnetic field on the solar surface forming active regions \citep[for a review, see \eg,][]{2014A&ARv..22...78W}. It is the motion of their photospheric footpoints that drive the evolution in the corona above \citep[\eg,][]{2014ApJ...787...87V,2019A&A...625A..53S}. Flares occur preferentially in regions that rapidly evolve and/or host complex magnetic field \citep[\eg, reviews by][]{2009AdSpR..43..739S,2009SSRv..144..351V}. Correspondingly, large amounts of electric current and magnetic helicity are induced in the solar atmosphere and the coronal magnetic energy increases. Magnetic helicity is a measure of the topological complexity of a magnetic field. In the presence of a flux rope it quantifies how much the flux rope is internally twisted, whereas in the presence of a sheared arcade, it quantifies how much it is sheared \citep[\eg, reviews by][]{2007AdSpR..39.1674D,2014SSRv..186..285P,2019LRSP...16....3T}. As it is tightly related to the structural complexity of the underlying magnetic field, the sign of magnetic helicity unambiguously relates to the sense of the twist and shear of the magnetic structure; namely, its right- or lefthanded nature.

Magnetic energy is released in parts in the process of magnetic reconnection during solar flares \citep[\eg, reviews by][]{2002A&ARv..10..313P,2011SSRv..159...19F} and is transformed into kinetic and thermal energy. In contrast, magnetic helicity is not dissipated on time scales relevant for solar flare processes \citep[\eg,][]{1984GApFD..30...79B} and is substantially reduced only when a complex magnetic field structure is removed from the corona; for instance, via a CME. One of the key challenges in solar physics today is to understand the physics of the magnetic field in solar active regions (ARs), both in space and time, and to use this understanding to predict upcoming flare/CME activity \citep[for a review see, \eg,][]{2018SSRv..214...46G}. 

At photospheric levels, parameters related to the magnetic field structure near the flare- and CME-associated (core) regions were found as indicative for upcoming flare activity \citep[\eg,][]{2007ApJ...655L.117S,2015ApJ...804L..28S}. At coronal levels, a clear correlation exists between the preflare coronal energy budget and upcoming flare activity, both in frequency and size \citep[][]{2014ApJ...788..150S}. A certain energy budget, however, does not seem to be indicative of whether a flare will occur; this is also because only a fraction of it is released during a flare \citep[\eg,][]{2012SoPh..276..133G}. This necessarily points at a limited indicative ability of the coronal energy content with respect to the upcoming flare activity of an AR. The numerical experiment carried out by \cite{2017A&A...601A.125P} showed that a certain preflare helicity budget alone is also not a sufficient criterion for flaring. The statistical analysis of \cite{2012ApJ...759L...4T} suggests that it is a combination of sufficient strength and complexity of the coronal magnetic field, which is related to enhanced flare productivity. As suggested from both, numerical experiments \citep[][]{2017A&A...601A.125P} and data-constrained modeling of selected ARs \citep{2019ApJ...887...64T}, larger preflare coronal budgets of magnetic energy and helicity do not necessarily result in larger flares to occur or flares of a preferred type to occur. Importantly, \cite{2017A&A...601A.125P} showed that relative measures of these coronal budgets are more successful in discriminating the flare potential of ARs, such as the energy ratio (\ie, the ratio of free to total energy) or the helicity ratio (\ie, the ratio of the helicity of the current-carrying field to the total helicity; for details see Sect.~\href{ss:modeling}{\ref{ss:modeling}}). This suggestion was soon thereafter confirmed via a corresponding pioneering data-based studies \citep[\eg,][]{2019ApJ...887...64T}.

While the prediction whether or not a flare will occur in an AR is already challenging, the prediction whether or not it will be associated to a CME is yet another story. Flares accompanied by a CME are referred to as eruptive, and confined otherwise \citep[\eg,][]{1986lasf.conf..332S}. The association rate of flares and CMEs steeply increases with flare size, from $\approx$\,60\% for flares of {\it Geostationary Operational Environmental Satellite} (GOES) class M3 or larger to $\approx$\,100\% for flares of GOES class X4 or larger \citep[][]{2005JGRA..11012S05Y}. In this context, it is important to understand what exactly determines the flare type (eruptive or confined). Observational aspects at photospheric levels considered in that context include the flare location within the host AR \citep[\eg,][]{2007ApJ...665.1428W,2018ApJ...853..105B} and the structural relationship between the flare site and host AR \citep[\eg,][]{2017ApJ...834...56T}. Today, it is generally accepted that it is in particular the interplay between the strapping field and the AR nonpotentiality that define flare confinement versus its eruptiveness in the form of a CME (see the statistical study of \cite{2022ApJ...926L..14L}, and the recent review by \cite{2022SoPh..297...59K}). Both a lower overall strength as well as a sufficiently strong decay rate with height are often found for ARs that are CME productive (\eg, \cite{2008ApJ...679L.151L} and \cite{2007ApJ...665.1428W}, respectively). Related theory suggests a magnetic flux rope to become unstable if it reaches a height (the so-called "critical" height for torus instability) at which the decay rate of the horizontal field exceeds a critical value \citep{2006PhRvL..96y5002K}. Dedicated data-based statistical studies indeed found corresponding differences in the magnetic field configurations prior to confined and eruptive flares in support of this hypothesis \citep[\eg,][]{2017ApJ...843L...9W,2018ApJ...853..105B,2020ApJ...900..128L}.

Given that only a fraction of an AR magnetic field is involved in the reconnection process \citep[\eg,][]{2017ApJ...845...49K,2018ApJ...853...41T,2020ApJ...900..128L}, it is important to combine measures that characterize the stability of the magnetic field in different parts of an AR. Measures involving the magnetic field of the entire AR (\ie, integrated over the corresponding coronal volume) include the coronal magnetic energy and helicity (ratio). Measures involving the magnetic field of the host AR on a photospheric level include, for instance, the photospheric helicity flux. A measure that involves the magnetic field in a small sub-volume of the AR, more precisely, the field associated to the flare polarity inversion line is the critical height for torus instability. Recent attempts have studied the accumulated helicity flux \citep{2022A&A...662A...6L} or instantaneous coronal helicity ratio \citep{2023A&A...volA..ppG} in combination with the critical height for torus instability and in relation to the CME-productivity of solar ARs. Similar findings were presented in such works, namely, that magnetic field configurations with a higher nonpotentiality are primarily associated  with magnetic structures with lower critical heights. Most importantly, the flare type was found to be considerably better segregated in terms of the critical height.

Here, we present an in-depth analysis of magnetic-field related quantities around large solar flares, that is, of GOES class M1 or larger. We study the time evolution of the magnetic field around 231 large flares to our knowledge the largest number of flares covered so far by an individual study dedicated to analyze the flare- and CME-associated coronal free magnetic energy and helicity budgets in a statistical way. Our main goal is to test whether these two quantities can provide us with insight into the flare type in terms of confined versus eruptive. Furthermore, we test the relation of these quantities with other eruptivity proxies used in literature, including the mean twist parameter \citep[e.g.,][]{2016ApJ...821..127B},  flux-$R$ measure \citep{2007ApJ...655L.117S}, and the critical height for torus instability \citep{2007ApJ...665.1428W}. Based on optimization-based nonlinear force-free (NLFF) magnetic field models we deduce magnetic energy and helicity budgets as well as related proxies (Sect.~\href{ss:modeling}{\ref{ss:modeling}}) to study the nonpotentiality of coronal (preflare) magnetic fields. We analyze the preflare coronal budgets in context with the type of upcoming flaring (confined or eruptive), and related this also to photospheric proxies of the AR's magnetic complexity used in the literature. Using the time profiles of the deduced coronal quantities, we inspect their long-term evolution (covering up to 48 hours around the flares) with respect to flare occurrences, applying superposed epoch analysis for this purpose (Sect.~\href{ss:sea_analysis}{\ref{ss:sea_analysis}}). The preflare time profiles of deduced coronal quantities are also used to inspect the relative similarity in time evolution, applying dynamic time warping for this purpose (Sect.~\href{ss:corr_analysis}{\ref{ss:corr_analysis}}). Our results are presented in Sect.~\href{s:results}{\ref{s:results}}, followed by an extended discussion in Sect.~\href{s:discussion}{\ref{s:discussion}} and a summary in Sect.~\href{s:summary}{\ref{s:summary}}.


\section{Data and methods}

\subsection{Flare sample}

We analyzed a total of 231 large flares (of GOES class M1 and above) that occurred during solar cycle 24 and  originated within 50\,degrees to the east or west of the central solar meridian (see Table\,\href{tab:flare_summary}{\ref{tab:flare_summary}} for details). Those 231 flares originated from 36 different ARs. For completeness, we note that we investigated 43 ARs during their disk passage in total, yet not all of them were included in the final analysis due to reasons summarized in the following. Six ARs have not been included in our sample since their spatial proximity
with neighboring ARs severely hampers the useful definition of a model field of view
(FOV); namely: those with the {\it National Oceanic and Atmospheric Administration} (NOAA) numbers 11613, 11877, 11884, and 11900, 11944, and 12173.  More precisely, in these cases significant strong magnetic flux is located near the boundaries of the ARs' FOV, a known disadvantage to any attempt of high-quality coronal magnetic field modeling \citep[\eg,][]{2009ApJ...696.1780D}. Our sample does also not cover NOAA~11402, as the corresponding magnetic field modeling was of insufficient quality. A detailed description of the magnetic field modeling and the related quality assessment is presented in Sect.~\href{ss:modeling}{\ref{ss:modeling}}.

The flare-AR associations were established by combining information of different online sources, including the SolarSoft Latest events database\footnote{\url{https://www.lmsal.com/solarsoft/latest_events/}\label{f:ssle}} and the Hinode Flare Catalog\footnote{\url{https://hinode.isee.nagoya-u.ac.jp/flare_catalogue/}\label{f:hfc}}. For a clarification of flare-CME associations the SOHO/LASCO CME catalog\footnote{\url{https://cdaw.gsfc.nasa.gov/CME_list/}\label{f:lasco}}, as well as the Solar Demon Flare and Dimming detection service\footnote{\url{https://www.sidc.be/solardemon/}\label{f:sde}} were used.

\subsection{Magnetic field modeling and helicity computation}\label{ss:modeling}

To study the coronal magnetic field configuration of solar ARs in space and time we used 
data products from  Helioseismic and Magnetic Imager \citep[HMI;][]{2012SoPh..275..229S} on board  {Solar Dynamics Observatory} \citep[SDO;][]{2012SoPh..275....3P}. To study the three-dimensional (3D) coronal magnetic field, we used {\sc hmi.sharp\_cea\_720s} data, providing the Lambert cylindrical equal area (CEA) projected photospheric magnetic field vector within automatically-identified AR 
patches \citep[][]{2014SoPh..289.3549B}. The spherical heliographic components, $B_r$, $B_\theta$, and $B_\phi$ \citep{1990SoPh..126...21G} relate to the heliographic field components as $[B_x, B_y, B_z]$\,=\,$[B_\phi, -B_\theta, B_r]$ \citep{2013arXiv1309.2392S}, where $x$, $y$, and $z$ indicate a direction pointing to the solar west, north, and vertically upwards, respectively. Around the time of registered large flares (GOES class M1 or larger) we used a 12-min cadence, and an one-hour cadence otherwise.

The spatial sampling of the CEA vector data is 0.03 CEA-degrees (corresponding to $\sim$\,360\,km at disk center). For further analysis, we binned the SHARP-CEA data by a factor of two. These data were used as input to a nonlinear force-free (NLFF) method in order to model the 3D coronal magnetic field in and around the ARs and to compute the instantaneous coronal budgets. We note here that the inclusion of "more real" information, in the form of observational data with a finer spatial sampling, does not fully guarantee a more internally consistent NLFF solution \citep[for a comparative study, see][]{2009ApJ...696.1780D}. In particular, \cite{2022A&A...662A...3T} demonstrated that for the optimization method an improved (finer) spatial sampling actually lowers the final NLFF model quality. Instead, the binning of the spatial sampling of the CEA input data by a factor of two was found to result in more satisfactory NLFF solutions, when using the relatively larger budgets of the coronal free energy and the helicity of the current-carrying magnetic field as indicators. It should be noted here that spatial-sampling-induced differences are relatively small compared to those arising from other sources of uncertainty, such as the usage of data from different instruments,  data calibration method used, and so on \citep[for  details, see Sect.~4 of][]{2022A&A...662A...3T}.

To perform the NLFF modeling, we applied the method of \cite{2012SoPh..281...37W} and combined the improved optimization scheme of \cite{2010A&A...516A.107W} and a multi-scale approach \citep[][]{2008JGRA..113.3S02W}. In our work, we applied a three-level multi-scale approach to preprocessed vector magnetic field data, the latter achieved by applying the method of \cite{2006SoPh..233..215W} to the photospheric vector magnetic field data (using the standard settings as suggested in \cite{2012SoPh..281...37W}). We performed the NLFF reconstruction using an enhanced weighting of the volume-integrated divergence ($w_d$\,=\,2 in Eq.~(4) of \cite{2012SoPh..281...37W}) to achieve solutions of sufficient solenoidal quality to be used as input to our helicity computation method (for a proof of concept, see the appendix of \cite{2009ApJ...696.1780D} and for a dedicated in-depth study, see \cite{2019ApJ...880L...6T}). For each of the individual NLFF solutions, a potential field was computed as a reference, with $\vBp$\,=\,$\nabla\phi$ being sufficient, where $\phi$ is the scalar potential subject to the constraint $\nabla_n\phi$\,=\,$\vB_n$ on $\partial\mathcal{V}$, where $n$ denotes the normal component with respect to the boundaries of $\mathcal V$. Based on those 3D magnetic fields, we defined the total magnetic energy as $\Et$\,=\,$\int_V$\,$B^2$\,$\dV$, the potential energy as $\Ep$\,=\,$\int_V$\,$B_0^2$\,$\dV$, and the free magnetic energy as $\Ef$\,=\,$\Et$\,$-$\,$\Ep$.

\begin{figure}[b]
\centering
\includegraphics[width=0.7\columnwidth]{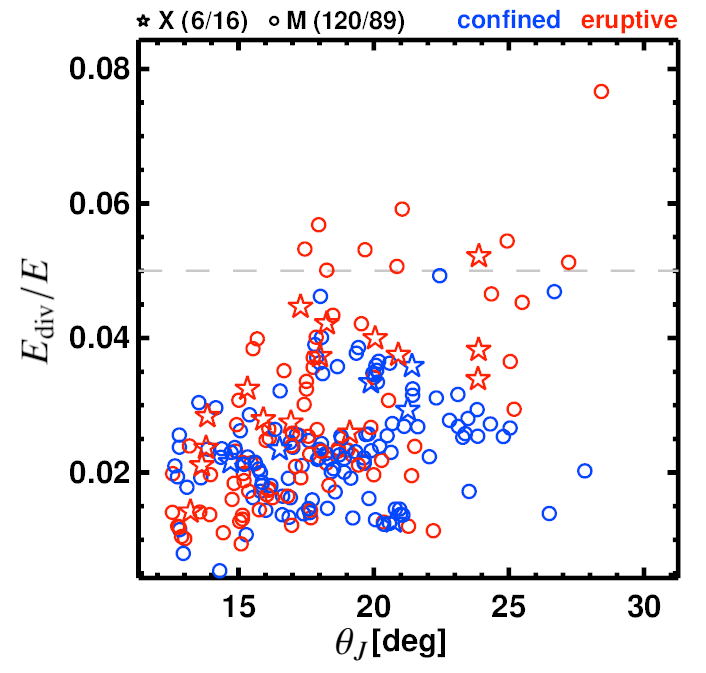}
\caption{Preflare NLFF model quality for the 231 large flares under study. The relative distribution of the solenoidal energy ratio ($\Edivprime$) vs.\ the current-weighted angle ($\thetaj$) is shown. Stars and circles represent X- and M-class flares, respectively. Numbers in brackets indicate the ratio of confined to eruptive flares within the tested sample. Blue (red) indicates an confined (eruptive) flare type.}
\label{fig:preflare_scatter_major_quality}
\end{figure}

Making advantage of the condition $\vB_n$\,=\,$\vBpn$ on $\partial\mathcal{V}$ and the additional constraint that the NLFF and potential field are solenoidal (\ie, $\divB$\,=\,0 and $\divBp$\,=\,0), we were able to compute the gauge-invariant relative helicity \citep[][]{1984JFM...147..133B,1985CPPCF...9...111F} as 
\begin{equation}
    \Hv = \int_V \left( \vA +\vAp \right) \cdot \left( \vB - \vBp \right) \dV,
\end{equation}
known to represent a meaningful quantity to compute and track the time evolution of a magnetic system within a finite model volume \citep{2012SoPh..278..347V}. In addition, we computed the separately gauge-invariant \citep[yet not individually conserved; see][]{2018ApJ...865...52L} contributions of the volume-threading and current-carrying field, following \cite{2003and..book..345B}, as
\begin{eqnarray}
    \Hpj &=& 2 \int_V \vAp \cdot \left( \vB - \vBp \right) \dV, \\
    \Hj &=& \int_V \left( \vA -\vAp \right) \cdot \left( \vB - \vBp \right) \dV,
\end{eqnarray}
respectively, in order to gain additional information on the time evolution of the respective nonpotential field, $\vBj$\,=\,$\vB$\,$-$\,$\vBp$. We obtained the vector potentials $\vA$ and $\vAp$ using the finite-volume (FV) method of \cite{2011SoPh..272..243T}. This method has been tested in a series of benchmark studies, where it was shown that it delivers helicity values in line with that of other tested methods for various numerical test setups \citep[][]{2016SSRv..201..147V} as well as for applications to real solar data \citep[][]{2021ApJ...922...41T}.

In addition to the above introduced volume-integrated quantities (also referred to as "extensive" measures), we compute specific relative ("intensive") measures. These include the (free) energy ratio ($\Efnt$), the helicity ratio ($\Hjn$), the flux-normalized helicity ($\Hjfn$) of the current-carrying field, and the flux-normalized total helicity ($\Hvfn$), where $\tilde\phi$ is half of the total unsigned photospheric flux, $\Fabs$ \citep[\eg,][]{2017A&A...601A.125P}. 


\begin{figure*}[ht]
\centering
\includegraphics[width=0.9\textwidth]{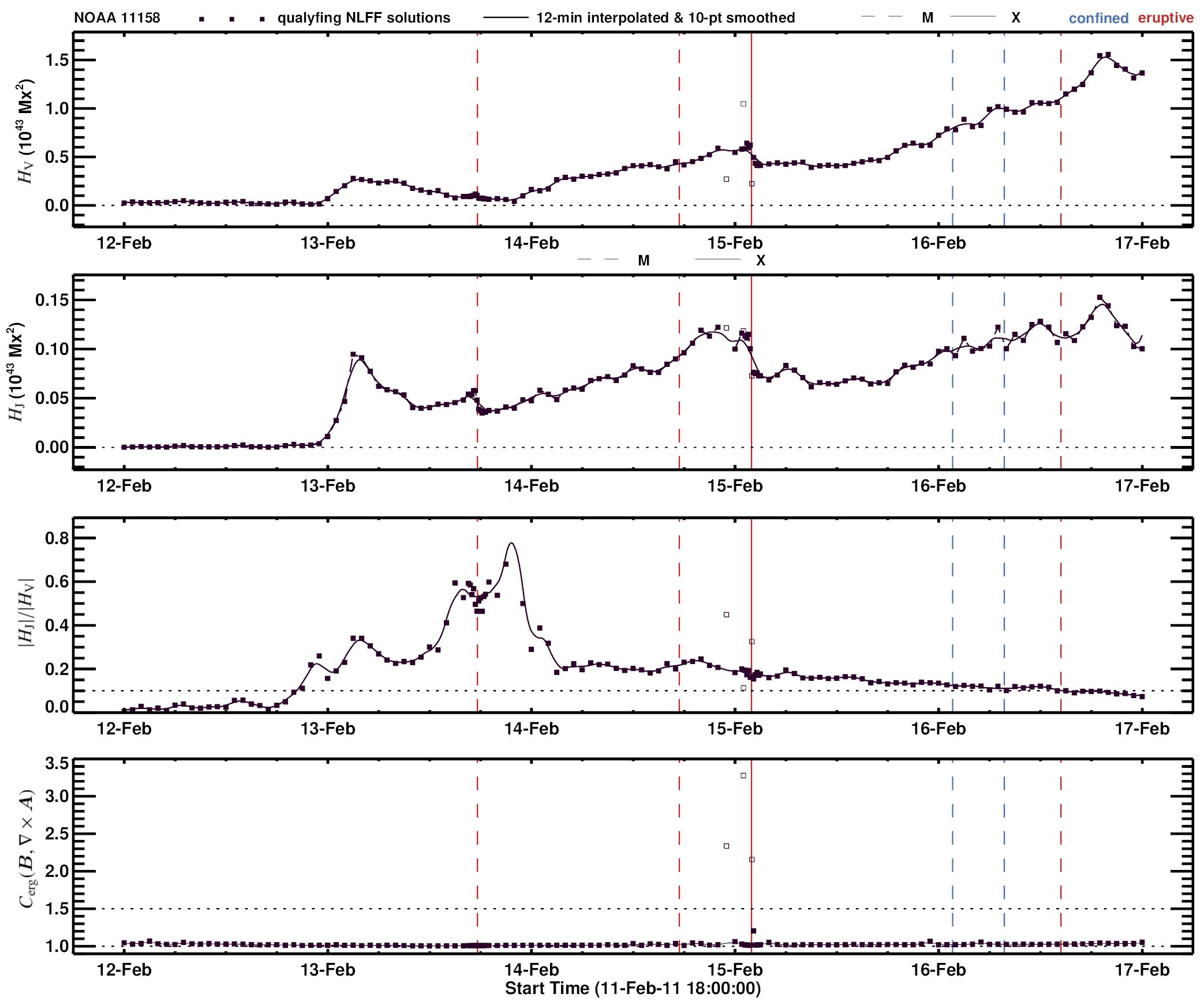}
\caption{Time evolution of selected quantities during the disk passage of NOAA~11158. From top to bottom: the total helicity ($\Hv$), helicity of the current-carrying field ($\Hj$), helicity ratio ($\Hjn$), and the energy correlation ($\cerg$) of the pair ($\vA$,$\nabla$\,$\times$\,$\vB$). (Dis-)Qualifying values are indicated by (empty) filled black squares. The black curve represents a fitting to the qualifying values onto a uniform (12-min) time cadence. Solid and dashed vertical lines mark the GOES start time of X- and M-class flares, respectively. Blue (red)  indicates an confined (eruptive) flare type.}
\label{fig:time_profile}
\end{figure*}

The NLFF solutions are required to fulfill certain conditions, above all, to be physically meaningful. This implies, for instance, $\Ep$\,$<$\,$\Et$, so that $\Ef$\,$>$\,0. In practice, nonphysical solutions are sometimes obtained when the coronal magnetic field is close to a potential configuration. Such NLFF models are excluded from our analysis. We define the NLFF modeling as successful and reliable once the following conditions are met. First, the NLFF models are expected to deliver a 3D corona-like model magnetic field with a vanishing Lorentz force and divergence. To quantify the force-free consistency we use the current-weighted angle between the modeled magnetic field and electric current density, $\thetaj$, \citep[][]{2006SoPh..235..161S}. Typically, $\thetaj$ is in the approximate range of 10$^\circ$ to 30$^\circ$. In only one case of our preflare event sample (SOL2014-12-14T19:25M1.6), the result is actually $\thetaj$\,$>$30$^\circ$ (see Fig.~\href{fig:preflare_scatter_major_quality}{\ref{fig:preflare_scatter_major_quality}}). Second, the energy contribution that arises from the nonsolenoidal component of the magnetic field, that is, from the finite divergence due to the necessarily limited numerical accuracy of the NLFF solution, $\Ediv$, has to be small \citep[for details, see the in-depth study of][]{2013A&A...553A..38V}. In the proof-of-concept study by \cite{2016SSRv..201..147V}, based on solar-like numerical experiments, it was suggested that a value of $\Edivprime$\,$\lesssim$\,0.08 is sufficient for a reliable helicity computation, reliable in the sense that the nonzero divergence of the underlying magnetic field model is not affecting the helicity computation. Correspondingly, we only considered NLFF solutions with values of $\Edivprime$\,$\leq$\,0.08. The follow-up study by \cite{2019ApJ...880L...6T} suggested an even lower threshold ($\Edivprime$\,$\lesssim$\,0.05) regarding solar applications. For completeness, we note here that for most of our preflare NLFF models (221 out of 231), we have: $\Edivprime$\,$<$$\,$\,0.05 (see dashed line in Fig.~\href{fig:preflare_scatter_major_quality}{\ref{fig:preflare_scatter_major_quality}} for reference). 

By definition, the vector potentials $\vA$ and $\vAp$ satisfy $\curlA$\,=\,$\vB$ and $\curlAp$\,=\,$\vBp$, respectively, where in our case $\vB$ and $\vBp$ denote the NLFF and potential magnetic field, respectively. Along with respecting $\divBp$\,=\,0 and using the Coulomb gauge ($\nabla$\,$\cdot$\,$\vA$\,=\,0 and $\nabla$\,$\cdot$\,$\vAp$\,=\,0), $\vA$ is retrieved by numerically solving a specifically designed Poisson problem, $\Delta\vA$\,=\,$-\vJ$, while $\vAp$ is retrieved by numerically solving a Laplace problem, $\Delta\vAp$\,=\,$0$, both subject to specifically designed consistent boundary conditions \citep[for details see][]{2011SoPh..272..243T}. To evaluate the internal consistency of the FV helicity method, we used the metrics introduced in Sect.~4 of \cite{2006SoPh..235..161S}. In particular, we considered  the energy correlation, $\cerg$, of the pair ($\vB$,$\curlA$). In this way, we determine how well the computed vector potential $\vA$ reproduces the energy contained in the input field $\vB$. Mathematically, it is defined as \begin{equation}
    \cerg=\frac{\sum_i|(\nabla\times\vA)_i|^2}{\sum_i|\vB_i|^2},
\end{equation} 
that is, as the ratio of the total magnetic energy in the field $\nabla$\,$\times$\,$\vA$ to the total magnetic energy in the NLFF solution ($\vB$). Therefore, $\cerg$ represents a global measure of the quality of the computed vector potential. 

To date it is unclear how a deviation from $\cerg$\,=\,1 relates to the precision of the deduced physical quantities. For a better understanding, we summarize some observed trends here. The magnetic energy is reproduced correctly to within $\approx$\,20\% (\ie, 0.8\,$\lesssim$\,$\cerg$\,$\lesssim$\,1.2) for the great majority of our NLFF models (for the example of NOAA~11158; see bottom panel in Fig.~\href{fig:time_profile}{\ref{fig:time_profile}}). For some ARs, a monotonic transition from smaller to larger (or the reverse) values is observed during disk passage. For some ARs the quality of the computed vector potentials is lower for the entire disk passage (NOAAs 11513, 12017, 12036, 12087, and 12242 in our AR sample). In most of those cases, the time profiles of the associated magnetic energies and helicities do not show obvious peculiarities. If so, we then used the underlying NLFF models for further analysis. Occasionally, we find pronounced sudden variations in the physical quantities computed from NLFF models when $\cerg$\,$\gtrsim$\,1.5. These noticeable variations {\sc{(i)}} do not occur in all of the employed physical quantities, {\sc{(ii)}} are differently pronounced for different quantities, and {\sc{(iii)}} their magnitude is not related to the degree of deviation of $\cerg$ from unity. To avoid erroneously interpreting such sudden and intense variations as  physical, we excluded such NLFF models manually from our analysis (marked by empty plot symbols in Figs.~\href{fig:time_profile}{\ref{fig:time_profile}}a--\href{fig:time_profile}{\ref{fig:time_profile}}c). 

Due to the filtering out of non-qualifying solutions based on the above requirements (and also since our model time cadence is 12-minutes around large flares, and one~hour otherwise) data gaps may arise in the time series of physical quantities. Therefore, we interpolated the time series of qualifying values to a uniform time cadence of 12~minutes throughout the entire disk passage and apply a Gaussian smoothing with a two-hour time window afterwards (see solid lines in Fig.~\href{fig:time_profile}{\ref{fig:time_profile}}). These interpolated time profiles are used for further analysis.

\subsection{Superposed epoch analysis} \label{ss:sea_analysis}

To perform a comparative study of the time evolution of the AR's coronal quantities, we applied a superposed epoch analysis (SEA), 
a statistical tool capable to identify existing patterns in time series and to study the statistical relationships between events of different type \citep{1913RSPTA.212...75C}. The method requires us to define the occurrence of target events (for instance a major flare) as key times within each considered time series (in our case given by the GOES start time). Then, the time series are shifted in time in order to align the event epochs with respect to the key times and sub-intervals around each key time were extracted. For convenience, we used the 12-min interpolated time series and extracted sub-intervals of equal length around each key time (the event epochs). Finally, the information contained in these synthesized event epochs was superposed and statistical properties were deduced. Using such an approach, recurring variations in the event epochs are assumed to be reinforced while random variations are assumed to cancel. For this work, we employed both mean and median values from the superposed epoch data, which allowed us to assess their statistical significance. In particular, we computed the lower and upper quartiles of the synthesized epoch data, as the 25th and 75th percentiles, respectively.

The SEA has been used in a number of solar flare studies so far. \cite{1975SoPh...41..227D} applied it in order to inspect enhanced flaring activity in context with the magnetic setting of the host AR as well as the properties of the large-scale (global) field. \cite{2015SoPh..290.2943S} used SEA to inspect the effect of major flares to the occurrence rate of energetic events in distant regions on the Sun. \cite{2010ApJ...723..634M} applied it to investigate the role of the complexity of the active-region magnetic field near the PIL during preflare times with respect to the properties of upcoming flaring. Only recently, \cite{2023ApJ...942...27L} used SEA to inspect the effect of X-class flaring onto the long-term coronal energy and helicity budgets. 

\begin{figure}[b]
\centering
\includegraphics[width=0.75\columnwidth]{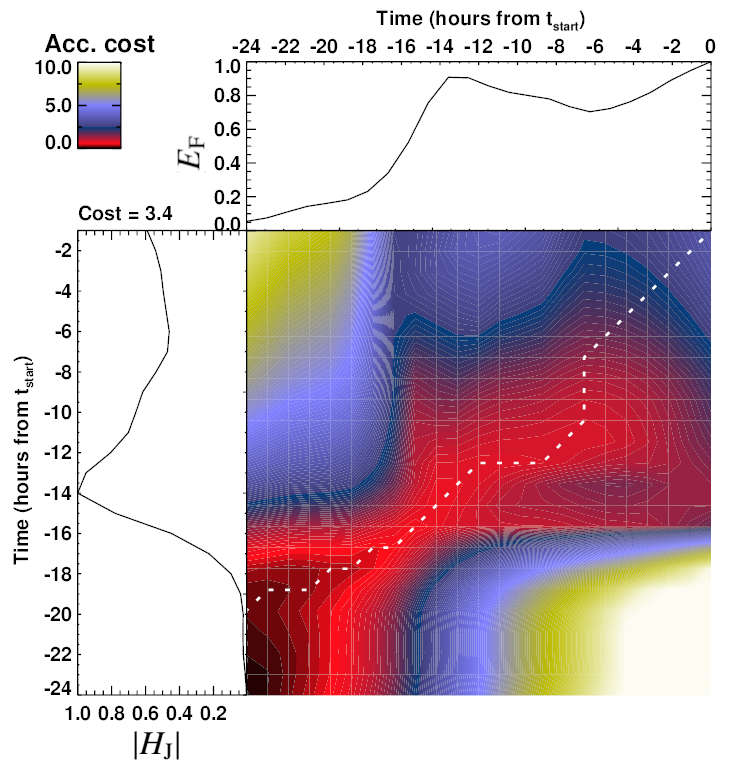}
\caption{Correlation analysis for the 24~hours leading up to the flare SOL2011-02-13T17:28M6.6. The normalized event epochs of the unsigned helicity of the current-carrying field ($\Hjabs$) and the free magnetic energy ($\Ef$) are shown in the vertically and horizontally oriented line profile, respectively. The accumulated cost matrix is shown color-coded. The white dotted line represents the optimal (warping) path.}
\label{fig:dtw_exemplary}
\end{figure}

\subsection{Correlation analysis} \label{ss:corr_analysis}

For the analysis of the time evolution of the various coronal quantities and to address how their time evolution relate to each other, we assume that the change to the coronal conditions due to the occurrence of large flare events are traceable in the form of distinct variations in the time profiles of the computed physical variables. We do not expect, however, the induced modulations of the time series to occur in a (strictly or nearly) synchronous way, to be of similar strength (\ie, a similar magnitude) or to affect similar portions of the time series (\ie, dissimilar time span), and so on. In other words, the time series of two different variables may show similar overall shapes but may not be exactly aligned in time. Yet is is exactly the shapes and trends which are of interest for us (rather than the absolute values), so that a suitable measure is needed to characterize them. 

\begin{figure*}
\centering
\includegraphics[width=0.9\textwidth]{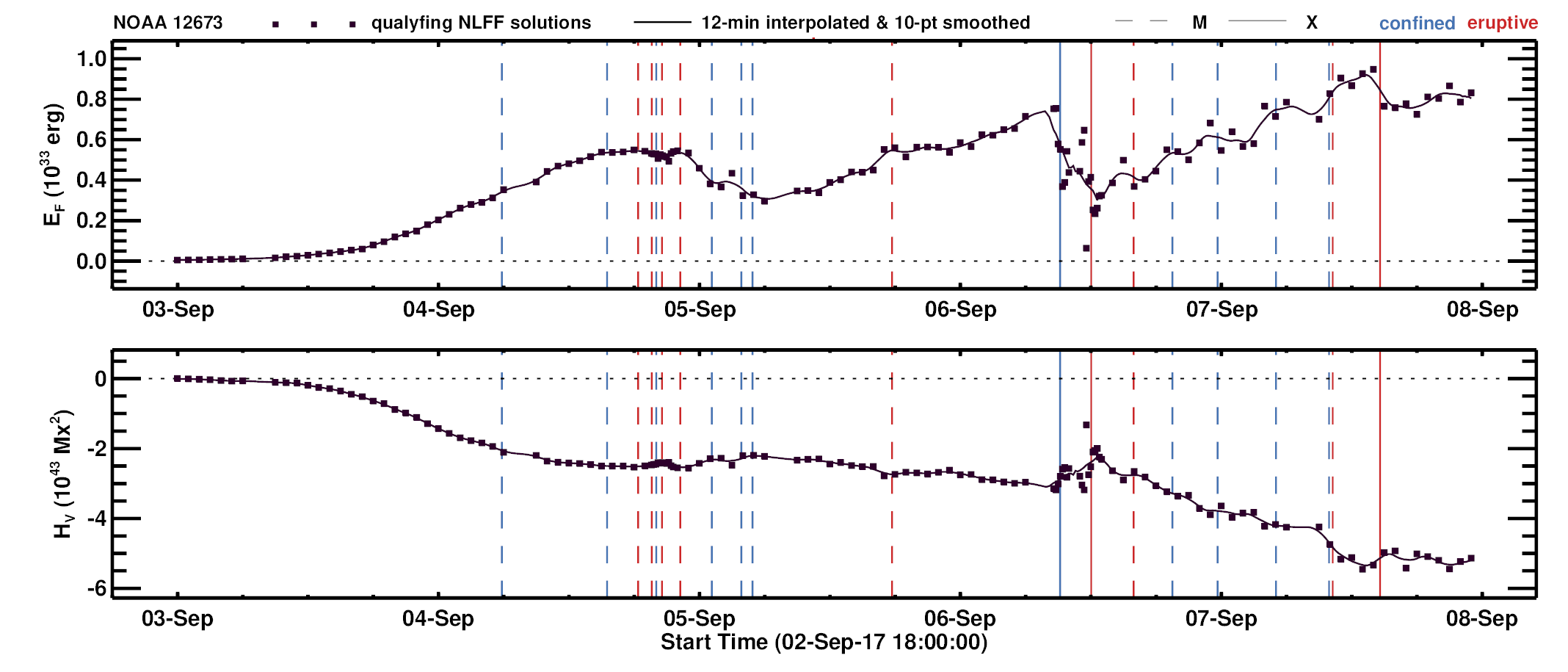}
\caption{Time evolution of coronal quantities during disk passage of NOAA 12673. (a) Free magnetic energy ($\Ef$) and (b) total helicity ($\Hv$). The black line represents a fitting to the values computed from qualifying NLFF models (squares) onto a uniform time cadence (12-minutes). Solid and dashed vertical lines mark the GOES start time of X- and M-class flares, respectively. Blue and red color indicate the flare type (confined and eruptive, respectively).}
\label{fig:time_profile_12673}
\end{figure*}

Dynamic time warping (DTW) is capable of doing so. Here, the agreement between two time series is quantified in the form of the cumulative cost ($\Ccum$), a single number that relates to the displacement between the two time series. The displacement is assessed in form of the so-called cost matrix (an example is shown in Fig.~\href{fig:dtw_exemplary}{\ref{fig:dtw_exemplary}}), computed based on the matrix of Euclidean distances constructed from a point-wise comparison of the values between all of the indices of the two time series. A so-called warping path is defined to characterize the mapping between the two time series in a way that (i) it starts and ends in the diagonally opposite corner cells of the cost matrix, (ii) it is continuous in the sense that it is only allowed to progress to adjacent (including diagonally adjacent) cells, and (iii) it is monotonous in the sense that it is monotonically spaced in time \citep[for further details see, \eg,][]{2001SIAM...1K}. Being faced with the fact that exponentially many warping paths satisfy these conditions, one seeks the path which minimizes the cumulative cost (indicated by the white dashed line Fig.~\href{fig:dtw_exemplary}{\ref{fig:dtw_exemplary}}). A smaller cumulative cost quite generally indicates a greater similarity between two time series. In this work, we use this classical form of DTW \citep[for a detailed discussion also with respect to alternative respective definitions see, \eg, Sect.~2.2 of][]{2022ApJ...927..187S}.

In the solar-flare context, \cite{2018ApJS..236...14M} involved DTW in an approach to the problem of CME-magnetic cloud association. We apply this technique here for the first time (to our knowledge) in order to assess the relative similarity of the time evolution of flare-related coronal quantities.

\section{Results and discussion}
\label{s:results}

\subsection{Long-term preflare time evolution}\label{ss:pre_long}

Pioneering simulation-based \citep{2017A&A...601A.125P} and data-constrained \citep{2019ApJ...887...64T} modeling of individual magnetic field configurations showed that some magnetic energy- and helicity-related measures exhibit characteristically different magnitudes for ARs that produce large confined or eruptive flares. The corresponding overall levels (magnitudes), however, persist during quite some time prior to the flaring and possibly already from the time of emergence onward. That means that the passing of a "critical" level in some of the measures is not indicative of when an (eruptive) flare will occur. To give an explicit example, Fig.~\href{fig:time_profile_12673}{\ref{fig:time_profile_12673}} shows the time evolution $\Ef$ and $\Hv$ during the disk passage of NOAA~12673. A confined X2.2 flare and an eruptive X9.3 flare originated from this AR on 2017 September 6 only about three hours apart in time (GOES start time 08:57~UT and 11:53~UT, respectively). Even larger preflare values of free energy and total helicity are estimated for the confined X2.2 flare, when compared to the preflare values of the following eruptive X9.3 flare \citep[for a dedicated case study see, \eg,][]{2019A&A...628A..50M}. 
In other words, when using the total budgets of magnetic energy and helicity as indicators, a (coronal) magnetic field owing to flare-favorable conditions ($\Hvabs$\,$\gtrsim$\,2\,$\times$\,$10^{42}$\,$\mxmx$ and $\Ef$\,$\gtrsim$\,4\,$\times$\,$10^{31}$\,$\erg$ as suggested in, \eg, \cite{2012ApJ...759L...4T}) do not necessarily produce an eruptive flare.

Recently, a first dedicated study by \cite{2023ApJ...942...27L} systematically examined the preflare time evolution of coronal quantities based on a sample of 21 X-class flares. Among others, they tested the overall budgets of magnetic energy and helicity on short ($\pm$\,five~hours around the flare time: see their Fig.~8) and longer time scales (from 48~hours before to 18~hours after each flare; see their Fig.~10). They found that in general the coronal budgets are similar prior to confined and eruptive flares. The authors based their analysis on synthesized information from all events via superposed epoch analysis (SEA). They explicitly note, however, that due to the low number of X-class flares suitable for study, no filtering regarding the possible occurrence of other flares within the studied time windows was done. In other words, the SEA-based time profiles might not purely represent the long-term preflare evolution, or might even be spurious. This is especially true for the small number of confined events in their study (six), out of which five originated from the same AR (12192). The authors suggested that one should expect stronger conclusions from SEA analysis when (i) applied to a larger sample of (especially confined) flares, and (ii) avoiding duplicate epochs in the superposition. 

Condition (ii) is a very restrictive one. It reduces our original event sample of 50 major flares (15/35 confined/eruptive) to 28 (5/23 confined/eruptive) and to 18 (1/17 confined/eruptive) events, when requiring a flare-less preflare time window of 12 and 24~hours, respectively. By flare-less we mean that no flare of GOES class M1 or larger occurred. Based on a such reduced event sample no meaningful analysis is possible, especially due to the class of confined flares being strongly under-represented. To weaken the effect of condition (ii) we therefore require no other major flare (GOES class M5 or larger) to happen within the preflare time window. Depending on the extent of the preflare time window (12 or 24~hours), 45 (15/30 confined/eruptive) or 37 (11/26 confined/eruptive) events remain for analysis. We chose the former of the two here, in order to include as many qualifying confined events as possible. 

\begin{figure*}
\centering
\includegraphics[width=0.75\textwidth]{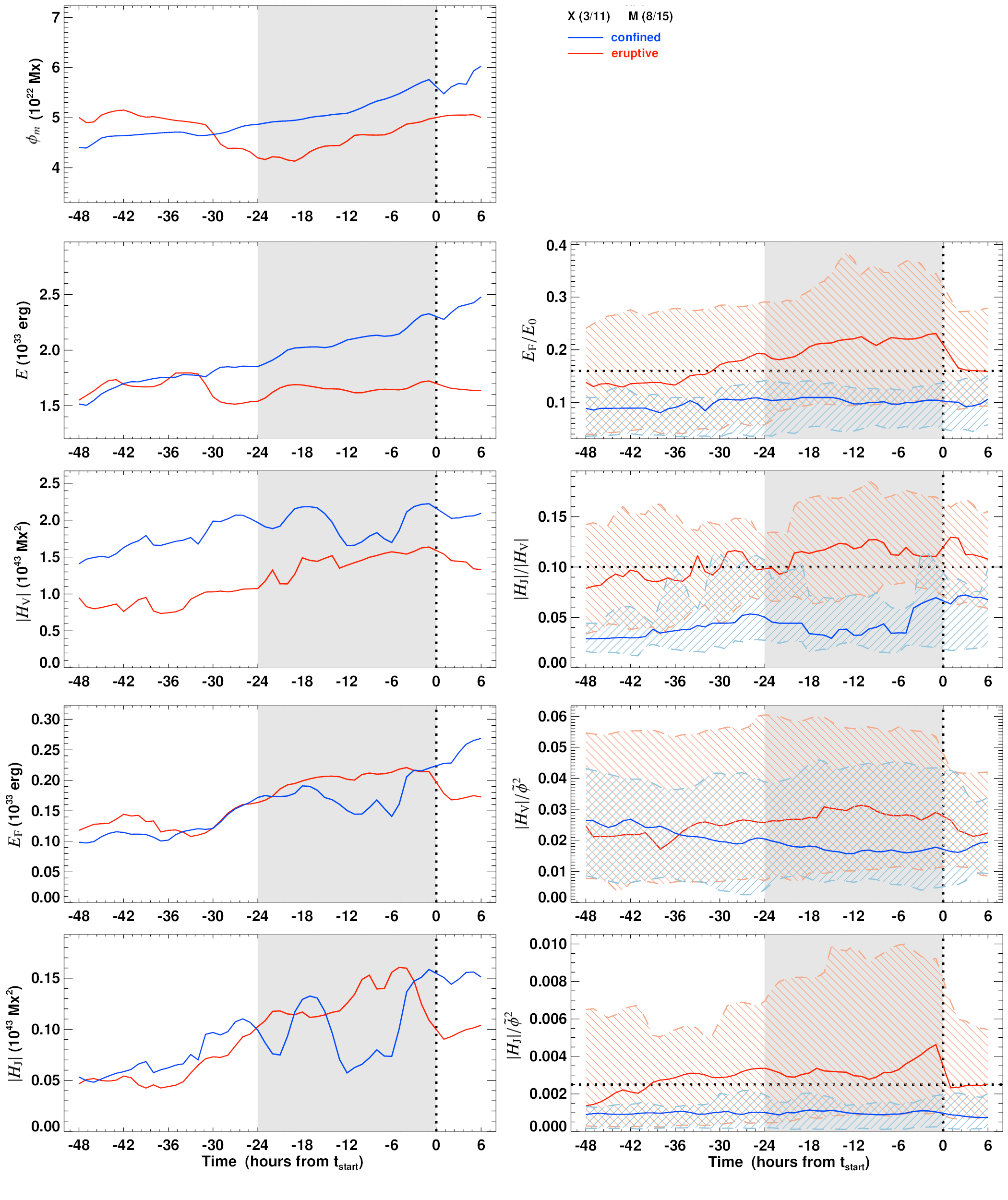}
\caption{Overall long-term preflare time evolution of selected coronal quantities deduced from superposed epoch analysis. Median values were computed separately for the subsets of confined (blue) and eruptive (red) flares. Covered is the time period from $-$48~hours to $+$6~hours around the flare start time (black dotted line at time $t_{\rm start}$\,=\,0). The 24~hours used for selection of qualified events is marked by the gray-shaded area. Numbers in brackets indicate the ratio of confined to eruptive flares within the tested sample. {\it Left column:} Superposed epoch data of (a) the total unsigned flux ($\Fabs$), (b) total magnetic energy ($\Et$), (d) total unsigned magnetic helicity ($\Hvabs$), (f) free energy ($\Ef$), and (h) unsigned helicity of the current-carrying field ($\Hjabs$). {\it Right column:} Superposed epoch data of the (c) free-to-potential magnetic energy ratio ($\Efnp$), (e) helicity ratio ($\Hjn$), (g) flux-normalized total helicity ($\Hvfn$), and (i) flux-normalized helicity of the current-carrying field ($\Hjfn$). Horizontal black dotted lines indicate respective critical values. The red and blue diagonally shaded regimes are bound by the upper and lower quartiles of the respective distributions.}
\label{fig:msea_major}
\end{figure*}

\begin{figure*}
\centering
\includegraphics[width=\textwidth]{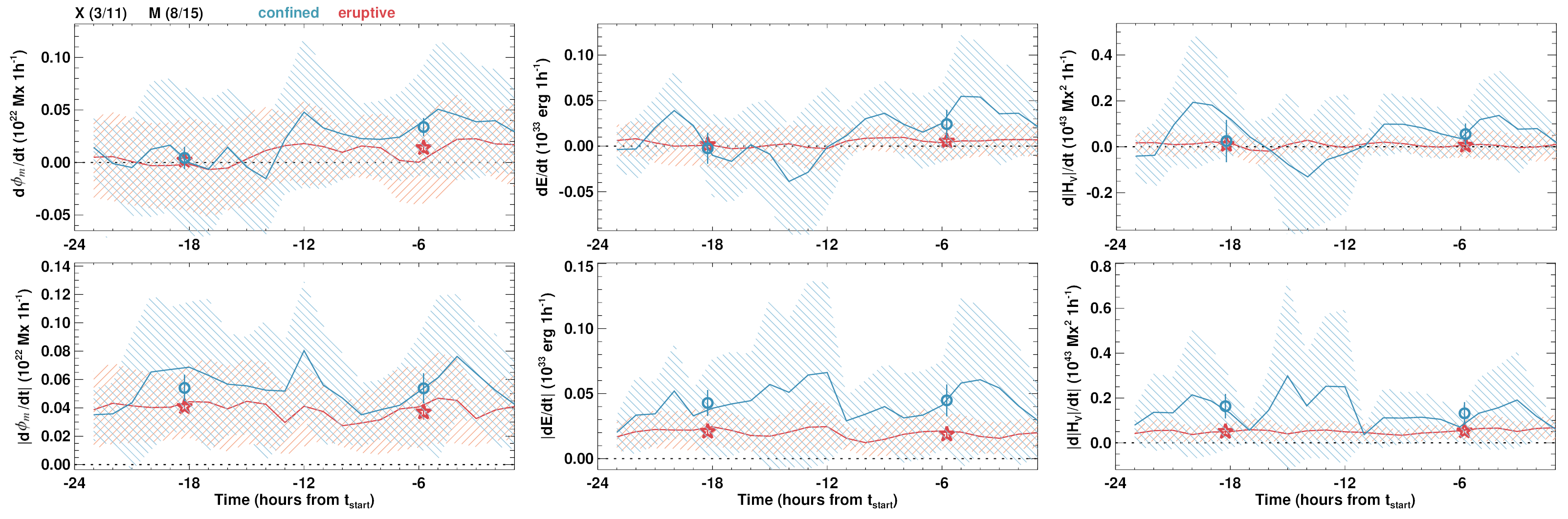}
\caption{Mean hourly variations (one-hour time gradient) in the overall preflare time evolution. Numbers in brackets indicate the ratio of confined to eruptive flares within the tested sample ($\Ssea$). Mean values for confined (blue) and eruptive (red) flares, computed from the synthesized preflare time profiles are shown. The shaded regimes indicate the standard deviation. Covered is the 24~hours time period prior to the flare start time ($t_{\rm start}$). The top panels show signed variations in the time evolution of the (a) unsigned magnetic flux ($\Fabs$), (b) total energy $\Et$, and (c) unsigned total helicity ($\Hvabs$). Positive/negative values indicate an/a respective increase/decrease over time. The bottom panels show the respective unsigned variations.}
\label{fig:grad1h_sea_major}
\end{figure*}

\subsubsection{Characteristic time scales and magnitudes}\label{sss:pre_long}

To be able to tell how unique (in time) it is that coronal conditions favor a specific type of flaring (confined or eruptive), we inspect overall trends obtained from the averaging of the superposed epoch data, separately for confined and eruptive major flares (11 and 26, respectively, out of our sample $\Smajor$; hereafter referred to as "$\Ssea$"). We note here that the spread for the extensive measures (left column of Fig.~\href{fig:msea_major}{\ref{fig:msea_major}}) is not shown in order to enhance visibility. This is because the spread is naturally very large as the values stem from ARs with very different magnetic properties, hence magnitudes of the computed extensive measures. The following trends are observed during the 24-hour flare-less preflare intervals (marked by gray shading) from these profiles. First, a monotonic increase of $\Fabs$ (an indication for ongoing flux emergence; see Fig.~\href{fig:msea_major}{\ref{fig:msea_major}}a) serves as a plausible explanation for the corresponding overall increases of the free energy and total helicity prior to eruptive flares (see red curves in Fig.~\href{fig:msea_major}{\ref{fig:msea_major}}f and \href{fig:msea_major}{\ref{fig:msea_major}}d, respectively). A similar argumentation for the time evolution prior to confined flares is hampered by the rather strong variations of the quantities (see blue curves in these panels). Second, the overall magnitudes of the coronal budgets appear somewhat higher prior to confined flaring than prior to eruptive flaring. The reason for this might simply be that three of the confined events originated from NOAA~12192 which, due to its exceptionally large spatial extent, hosted an exceptionally large amount of magnetic flux \citep[for dedicated in-depth studies see, \eg,][]{2015ApJ...801L..23T,2015ApJ...804L..28S}. As a consequence, the coronal energy and helicity budgets were about a factor ten larger compared to that of "typical" solar ARs \citep[for a comparative study see, \eg,][]{2019ApJ...887...64T}, necessarily resulting in larger mean and median values upon averaging over all of the confined events. Third, the overall time evolution of the extensive measures prior to confined and eruptive flaring appears rather similar. Here, our refined analysis --- considering a larger number of flares and requiring a 24-hour preflare time interval to be free of major flare occurrence --- strengthens the earlier findings of \cite{2023ApJ...942...27L}.

Going beyond this earlier study by \cite{2023ApJ...942...27L},  in the present work, the preflare evolution of the relative (intensive) measures has been analyzed in such a systematic way for the first time  (right column in Fig.~\href{fig:msea_major}{\ref{fig:msea_major}}). Since the employment of relative measures compensates the possibly existing major differences in the underlying magnetic settings, this representation allows us to state with confidence that the coronal conditions prior to eruptive and confined flaring are characteristically different: the average values of the intensive proxies (both mean and median) are considerably larger prior to eruptive flaring. Most pronounced variations over time are observed for $\Hjn$, which results from the strong sensitivity of the underlying extensive measures ($\Hj$ and $\Hv$) to the degree of nonpotentiality in the  magnetic field. The median time evolutions of $\Hvfn$ and $\Hjfn$ seem rather similar, so that use only $\Hjfn$ in the considerations hereafter. Typical values characterizing the corona prior to eruptive flares appear as $\Efnp$\,$\gtrsim$\,0.16, $\Hjn$\,$\gtrsim$\,0.1, and $\Hjfn$\,$\gtrsim$\,2.5\,$\times$\,$10^{-3}$ (see horizontal dotted lines the respective panels for reference). In comparison, the corresponding typical values prior to confined flaring reside clearly below these threshold values (indicated in blue color). These values seem to segregate most suitable the mean and median values of the given distributions during the major-flare less 24-hour preflare period and will be referred to as "critical values" hereafter. 

Most importantly, it is to be noticed that favorable conditions for eruptivity (using the critical values of $\Efnp$, $\Hjn$, and $\Hjfn$ defined above) are met during at least 12 to 24~hours prior to flare onset and possibly for much longer than that (see Fig.~\href{fig:msea_major}{\ref{fig:msea_major}}c, \href{fig:msea_major}{\ref{fig:msea_major}}e, and \href{fig:msea_major}{\ref{fig:msea_major}}i, respectively). The significant overlap of the flare-type related distributions (indicated by the shaded regimes, bound by the upper and lower quartiles) suggests that these measures are of limited predictive power in an individual case-by-case sense. It may still be, however, that the coronal quantities show time variations (change rates) on characteristically different time scales or of characteristically different magnitude prior to eruptive or confined flaring. Therefore, in Sect.~\href{sss:grad}{\ref{sss:grad}} we inspect the time derivatives of the superposed epoch data. It may also be that individual quantities evolve in a more or less similar manner with respect to other quantities prior to confined or eruptive flares. To test this aspect, in Sect.~\href{sss:dtw}{\ref{sss:dtw}} we applied DTW to selected pairs of variables.

\subsubsection{Characteristic change rates}\label{sss:grad}

In the following, we describe the trends observed for the change rate computed from the superposed epoch data over a full day prior to the occurrence of major flares. 
The change rate of the total unsigned flux, total energy and total unsigned helicity are on overall positive, especially during the last 12~hours prior to flare start (see top row in Fig.~\href{fig:grad1h_sea_major}{\ref{fig:grad1h_sea_major}}), in accordance with overall trends in the superposed epoch data discussed in Sect.~\href{sss:pre_long}{\ref{sss:pre_long}} (and compare Fig.~\href{fig:msea_major}{\ref{fig:msea_major}}). Noteworthy, the change rates of the budgets appear somewhat larger prior to confined flaring. On the other hand, the change rates prior to eruptive flaring appear more smooth. The unsigned change rates exhibit similar magnitudes prior to confined and eruptive flaring (see bottom panels in Fig.~\href{fig:grad1h_sea_major}{\ref{fig:grad1h_sea_major}}). 

We find that typical modulations to the coronal budgets in the absence of major flaring and irrespective of the type of upcoming flaring are $\langle{|\rm d}\Fabs|\rangle$\,$\approx$\,4.4\,$\pm$\,0.5\,$\times$\,$10^{20}$\,\mx/h, $\langle{|\rm d}\Et|\rangle$\,$\approx$\,2.7\,$\pm$\,0.4\,$\times$\,$10^{31}$\,\erg/h, and that of the free magnetic energy as $\langle{|\rm d}\Ef|\rangle$\,$\approx$\,0.91\,$\pm$\,0.1\,$\times$\,$10^{31}$\,\erg/h. For the coronal helicities we find characteristic changes of $\langle{|\rm d}\Hv|\rangle$\,$\approx$\,7.8\,$\pm$\,1.6\,$\times$\,$10^{41}$\,$\mxmx$/h and $\langle{|\rm d}\Hj|\rangle$\,$\approx$\,1.0\,$\pm$\,0.2\,$\times$\,$10^{41}$\,$\mxmx$/h.

\begin{figure*}
\centering
\includegraphics[width=0.9\textwidth]{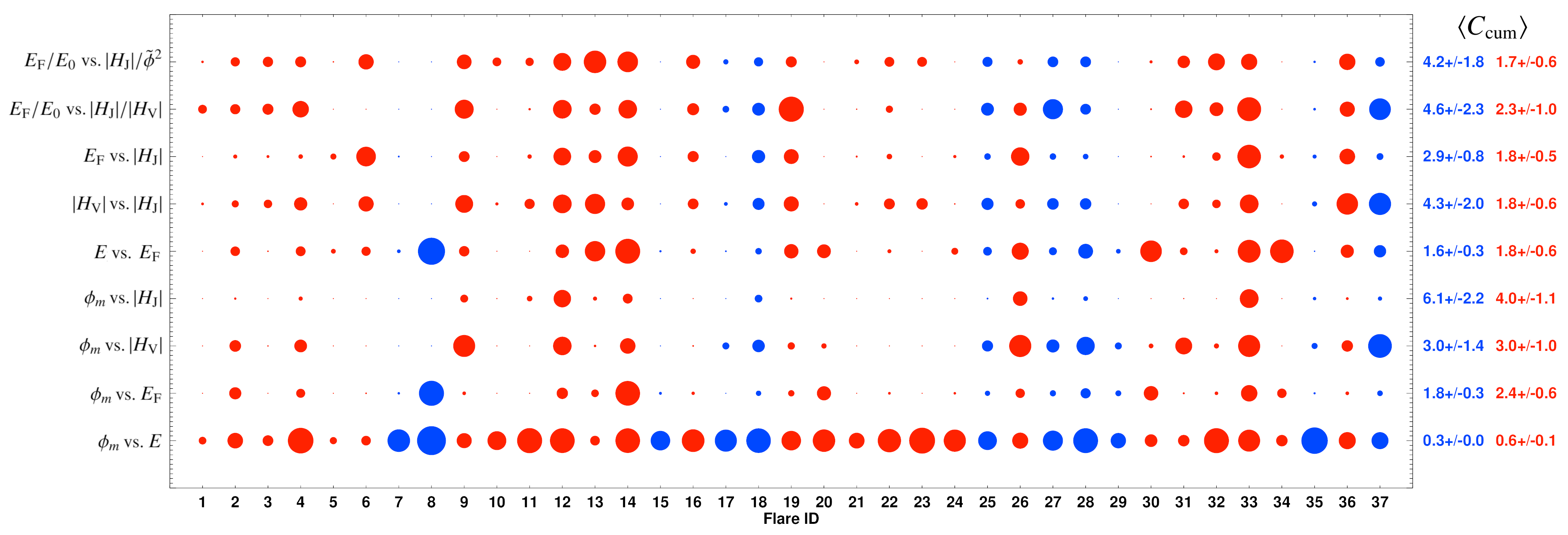}
\caption{Cumulative cost of pairs of variables ($\Ccum$; shown as bullets), computed for the normalized preflare time profiles during the 24~hours leading to the flares of sample $\Ssea$ (flare ID is shown along of the x-axis and associated to specific flare events in the last column of Table\,\href{tab:flare_summary}{\ref{tab:flare_summary}}). The size of the plot symbols (bullets) relate to the inverse of $\Ccum$, \ie, larger symbols indicate a larger similarity of two time profiles. Respective mean values, $\mCcum$, computed across all flares of a certain type are given to the right of the figure. Red and blue color indicate eruptive and confined flares, respectively.}
\label{fig:dtw_major}
\end{figure*}

\subsubsection{Correlation aspects}\label{sss:dtw}

Next, we sought to establish a quantification of the preflare time evolution of different extensive and intensive physical quantities with respect to each other. First, we aimed to get an overview of the general behavior prior to major flares. To do so, we applied DTW (for details see Sect.~\href{ss:corr_analysis}{\ref{ss:corr_analysis}}) to the 24~hours leading to the individual flares. We note again that no other major flares occurred during this preflare time period. To compensate for the known different overall magnitudes of the compared physical quantities, all time profiles were normalized, for which the respective maximum values during the preflare time window were used. 
Then, we assessed the cumulative cost, $\Ccum$, for selected pairs of variables for each flare of the sample, $\Ssea$, and computed the mean value across all flares ($\mCcum$; listed in Table\,\href{table:preflare_cost}{\ref{table:preflare_cost}}).

Overall (i.e., neglecting the type of upcoming flaring), the smallest value of $\mCcum$ during the 24~hours leading to a major flare is found for the pairs ($\Fabs$,$\Et$) (see Table~\href{table:preflare_cost}{\ref{table:preflare_cost}}). Together with $\mCcum$\,=\,0.44$\pm$0.14 for the pair ($\Fabs$,$\Ep$) this implies that the photospheric flux is strongly conditioning the coronal energy budget. In contrast, for the conditioning of the nonpotential energy also other sources are important, as the mean cumulative cost for the pair ($\Fabs$,$\Ef$) is considerably larger ($\mCcum$\,=\,2.32$\pm$0.59). The time evolution of $\Ef$ is more closely related to that of $\Et$ and $\Hvabs$ ($\mCcum$\,$\lesssim$\,2) than to that of the helicity of the current-carrying field ($\mCcum$\,=\,2.57$\pm$1.13), though the significance of these trends is certainly to be viewed in context with the rather large uncertainties. Largest cumulative costs are found for the pairs ($\Fabs$,$\Hjabs$), ($\Ep$,$\Hjabs$), and ($\Et$,$\Hjabs$) indicating that flux emergence is of little importance for the time evolution of the nonpotential energy.

Next, we addressed whether the overall similarity in the preflare time evolution of selected pairs of variables is indicative of the type of upcoming flaring. In Fig.~\href{fig:dtw_major}{\ref{fig:dtw_major}}, $\Ccum$ for different pairs of variables is shown for each analyzed flare (in the form of bullets and with the size of the plot symbols being inversely proportional to $\Ccum$, that is, being larger for a better match of two time profiles). Obviously for some flares, the majority of tested pairs evolve much more similar with respect to each other (e.g., flares no.\ 14 and 33, where $\Ccum$ is noticeable small) than in other flares. In contrast, strongly dissimilar time evolution of the majority of tested pairs is found (\ie, large values of $\Ccum$; for example, flares no.\ 1, 5 and 21). This is actually an interesting aspect and it is left for a future study to relate the similarity of the time evolution of different coronal quantities to the properties of the host AR. Rather independent of that, however, seems that for all of the flare events strongest similarities are found for the time evolution of the unsigned flux ($\Fabs$) and the coronal energies $\Et$ (and $\Ep$, though not shown explicitly), in line with the overall trends deduced from Table~\href{table:preflare_cost}{\ref{table:preflare_cost}}.

\begin{table}
\scriptsize
\setlength{\tabcolsep}{2.5pt}
\caption{
Relative similarity of the pre-flare time evolution of photospheric and coronal quantities.
} 
\label{table:preflare_cost}
\centering
\begin{tabularx}{\columnwidth} { 
  | >{\raggedright\arraybackslash}X 
  | >{\centering\arraybackslash}X 
  | >{\centering\arraybackslash}X 
  | >{\centering\arraybackslash}X 
  | >{\centering\arraybackslash}X 
  | >{\raggedleft\arraybackslash}X | 
  }
\hline
~ & $\Fabs$  & $\Et$ & $\Ef$ & $\Hvabs$ & $\Hjabs$ \\ 
\hline
$\Fabs$ & $\times$ & \scol 0.53$\pm$0.15 & \mcol 2.32$\pm$0.59 & \mcol 3.03$\pm$1.18 & \wcol 4.65$\pm$1.51 \\
\hline
$\Et$ & ~ & $\times$ & \scol 1.77$\pm$0.56 & \mcol 2.60$\pm$1.06 & \mcol 4.11$\pm$1.40 \\
\hline
$\Ef$ & ~ & ~ & $\times$ & \scol 1.82$\pm$0.64 & \mcol 2.57$\pm$1.13 \\
\hline
$\Hvabs$ & ~ & ~ & ~ & $\times$ & \scol 2.19$\pm$0.61 \\
\hline
$\Hjabs$ & ~ & ~ & ~ & ~ & $\times$ \\
\hline
\end{tabularx}
~\\~\\
\begin{tabularx}{\columnwidth} { 
  | >{\raggedright\arraybackslash}X 
  | >{\centering\arraybackslash}X 
  | >{\centering\arraybackslash}X 
  | >{\raggedleft\arraybackslash}X | 
  }
\hline
~ & $\Efnp$ & $\Hjn$ & $\Hjfn$\\ 
\hline 
$\Efnp$ & $\times$ & \mcol 3.06$\pm$1.40 & \mcol 2.52$\pm$1.08\\
\hline
$\Hjn$ & ~ & $\times$ & \scol 1.98$\pm$0.77 \\
\hline
$\Hjfn$ & ~ & ~ & $\times$ \\ 
\hline
\end{tabularx}
\tablefoot{
Average values are computed from the cumulative costs of the  flares of sample $\Ssea$. Uncertainties are defined based on the respective standard deviations. The cell colors are indicating the degree of similarity, and refer to $\mCcum$\,$\leq$\,2.2 (dark red) as "highly similar", 2.2\,$<$\,$\mCcum$\,$\leq$\,4.4 (orange) as "moderately similar", and 4.4\,$<$\,$\mCcum$ (light red) as "little similar". 
}
\end{table}

Intuitively, based on the comparison of the preflare time evolution of, for instance, $\Fabs$ and $\Et$ in Fig.~\href{fig:msea_major}{\ref{fig:msea_major}}, we would actually expect the largest overall similarity in the time evolution independent of the type of upcoming flaring, and for which indeed lowest values of $\mCcum$ are found (indicated to the right of Fig.~\href{fig:dtw_major}{\ref{fig:dtw_major}}, separately for the subsets of confined and eruptive flares). However, we also find flare-type specific trends. For instance, some of the extensive measures evolve in a more similar way with respect to each other prior to confined flaring. These include ($\Fabs$,$\Et$) and ($\Fabs$,$\Ef$) for which we find $\mCcum$\,=\,0.3\,$\pm$\,0.0 and 1.8\,$\pm$\,0.3, respectively, i.e., smaller values than prior to eruptive flaring ($\mCcum$\,=\,0.6\,$\pm$\,0.1 and 2.4\,$\pm$\,0.6, respectively). In contrast, the pairs ($\Fabs$,$\Hjabs$) and ($\Ef$,$\Hjabs$), for example, evolve more similar prior to eruptive flaring ($\mCcum$\,=\,4.0\,$\pm$\,1.1 and 1.8\,$\pm$\,0.5, respectively) than prior to confined flaring ($\mCcum$\,=\,6.1\,$\pm$\,2.2 and 2.9\,$\pm$\,0.8, respectively). Interestingly, the relative (intensive) proxies ($\Efnp$, $\Hjn$, and $\Hjfn$) evolve much more similar in time prior to eruptive flares (($\mCcum$\,$\lesssim$\,2) than prior to confined flares ($\mCcum$\,$\gtrsim$\,4).

\begin{figure*}
\centering
\includegraphics[width=0.9\textwidth]{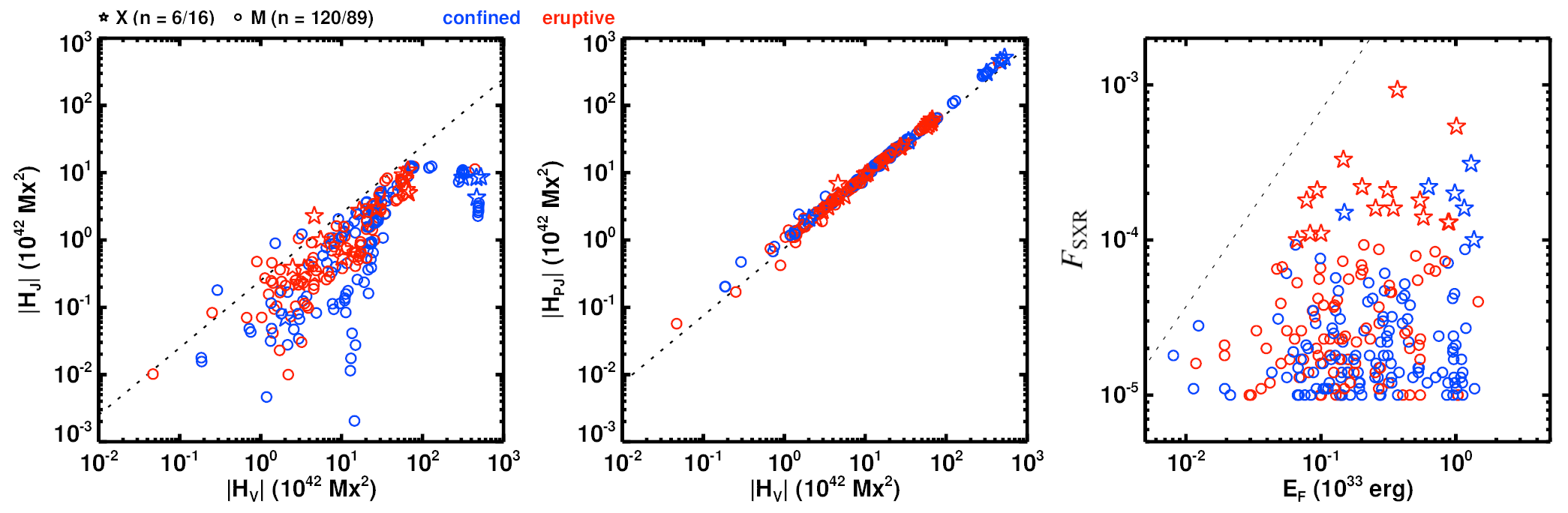}
\caption{Preflare values of selected measures for 231 flares. Numbers in brackets indicate the ratio of confined to eruptive flares within the tested sample $\Sall$. (a) Unsigned helicity of the current-carrying field vs.\ unsigned total helicity. The dotted line marks the $\Hjabs$/$\Hvabs$\,=\,0.25 level. (b)  Unsigned volume-threading helicity vs.\ unsigned total helicity. The dotted line marks the $|\Hpj|/\Hvabs$\,=\,1.0 level. (c) GOES 1--8\,\AA peak flux ($F_{\rm SXR}$) vs.\ free energy. The dotted line indicates the bolometric flare energy, using the relation $\log(E_{\rm bol})$\,=\,0.79$\pm$0.1\,$\log(F_{\rm SXR})$\,$+$\,34.5$\pm$0.5 \citep{2011A&A...530A..84K}. Blue and red color indicates confined and eruptive flare type, respectively. X-class and M-class flares are shown as stars and circles, respectively.}
\label{fig:preflare_scatter}
\end{figure*}

\subsection{Immediate preflare conditions}
\label{ss:pre}

We found favorable conditions for large flaring as $\Hv$\,$\gtrsim$\,1\,$\times$\,$10^{42}$\,$\mxmx$ and $\Ef$\,$\gtrsim$\,2\,$\times$\,$10^{31}$\,$\erg$ (see Fig.~\href{fig:preflare_scatter}{\ref{fig:preflare_scatter}}a and \href{fig:preflare_scatter}{\ref{fig:preflare_scatter}}c, respectively). Our analysis of a statistically meaningful number of large flares 231 flares of GOES class M1 or larger) therefore supports the earlier results presented in the statistical study of \cite{2012ApJ...759L...4T}, comprising 162 events (see their Fig.~2). These authors suggested large flares to occur in ARs with a free energy exceeding 4\,$\times$\,$10^{31}$\,$\erg$ and a relative helicity exceeding 2\,$\times$\,$10^{42}$\,$\mxmx$, the latter agreeing well with estimates of helicity contents of typical CMEs. For comparison, \cite{2022A&A...662A...6L} suggested thresholds for an AR to become CME-productive as $\Hv$\,$\gtrsim$\,9\,$\times$\,$10^{41}$\,$\mxmx$ and $\Et$\,$\gtrsim$\,2\,$\times$\,$10^{32}$\,$\erg$, based on the analysis of photospheric helicity and energy fluxes in newly emerging ARs. 

We identified some specifics in the distributions of $\Hj$ and $\Hpj$ with respect to the total helicity budget. We find that $\Hjabs$\,$<$\,0.25\,$\cdot$\,$\Hvabs$ in 96.5\% prior to the studied flare events (cf.\ dashed line in Fig.~\href{fig:preflare_scatter}{\ref{fig:preflare_scatter}}a), indicating that in ARs productive of large flares, the contribution of the current-carrying field to the total coronal helicity budget rarely exceeds 25\%. Closer inspection also reveals that for smaller preflare total helicity budgets (say, $\Hvabs$\,$\lesssim$\,5\,$\times$\,$10^{42}$\,$\mxmx$), the contribution of the current-carrying field, $\Hj$, tends to be smaller prior to confined flaring. This indicates that eruptive flaring might require a certain level of nonpotentiality in the underlying magnetic field. Interestingly, for values of $\Hvabs$\,$\gtrsim$\,$10^{44}$\,$\mxmx$, a plateau is noticeable in the distribution of $\Hjabs$ (saturating at values of $\Hjabs$\,$\lesssim$\,2\,$\times$\,$10^{43}$\,$\mxmx$; see Fig.~\href{fig:preflare_scatter}{\ref{fig:preflare_scatter}}a), while the unsigned volume-threading helicity, $|\Hpj|$, shows an ongoing increase for increasing values of $\Hvabs$ (see Fig.~\href{fig:preflare_scatter}{\ref{fig:preflare_scatter}}b). This may indicate that there exists some limit for the nonpotentiality of a current-carrying structure that a solar AR can host. However, it should be considered that the data points in the regime $\Hv$\,$\gtrsim$\,2\,$\times$\,$10^{44}$\,$\mxmx$ are associated to flares hosted by a single AR (NOAA~12192) which is known as the solar AR with an exceptionally large magnetic flux ($\gtrsim$\,1.5\,$\times$\,$10^{23}$\,$\mx$; for dedicated studies, see, \eg, \cite{2015ApJ...801L..23T,2020ApJ...900..128L}). 
Figure~\href{fig:preflare_scatter}{\ref{fig:preflare_scatter}}c shows that major confined flares are associated to ARs which host a larger free energy budget, compared to ARs hosting major eruptive flares. In this context, we also note that \cite{2020ApJ...900..128L} have shown that large flares that are produced by ARs containing large magnetic fluxes are mostly confined. However, in general the study of a larger sample of large confined flares (produced by different ARs) is desirable to further substantiate these findings.

\begin{figure}
\centering
\includegraphics[width=\columnwidth]{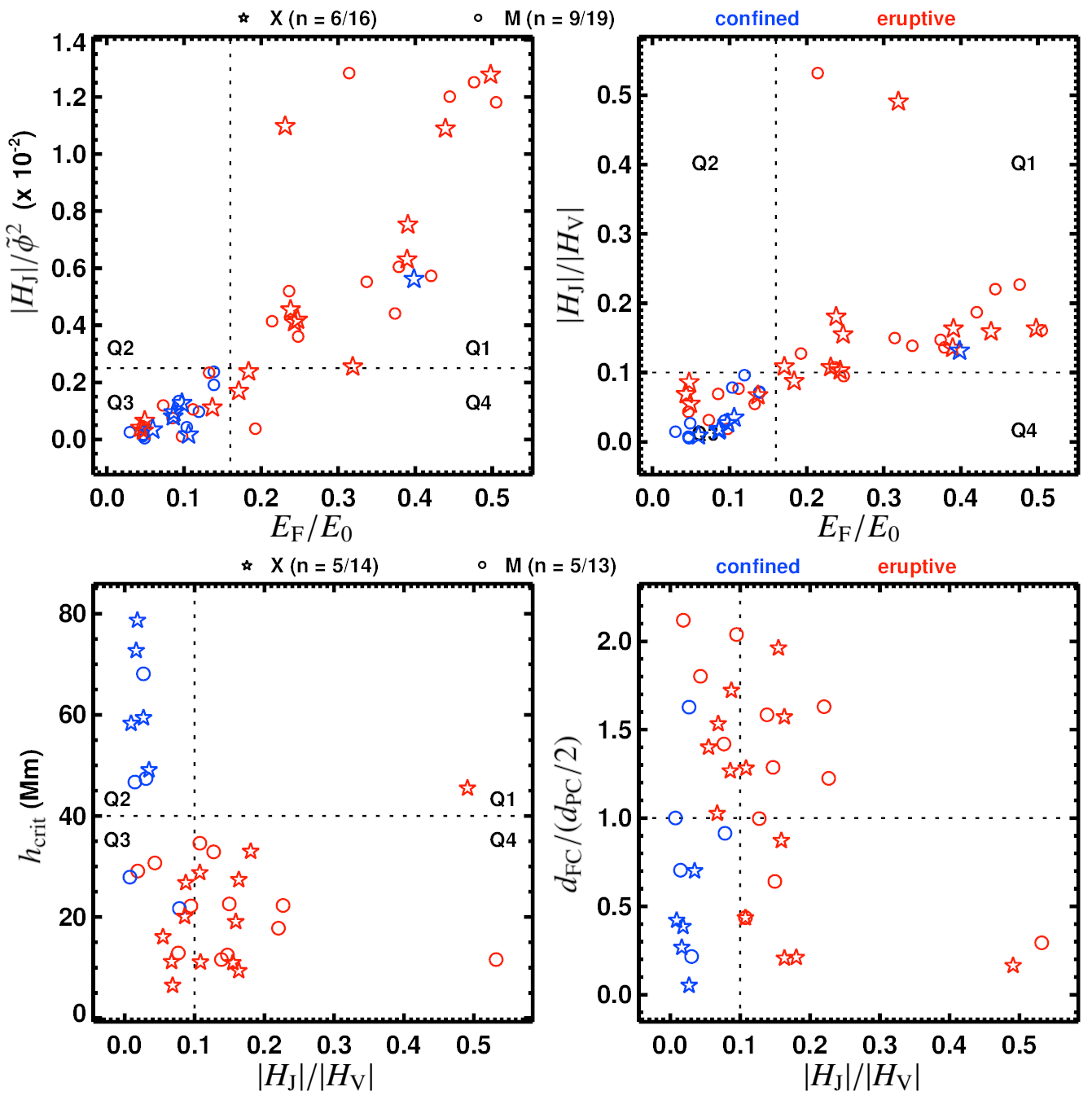}
\caption{Preflare values of selected measures. (a) Flux-normalized helicity of the current-carrying field vs.\ free energy ratio, and (b) helicity ratio vs.\ free energy ratio, for the event sample $\Smajor$. (c) Critical height for torus instability vs.\ helicity and (d) normalized flare distance vs.\ helicity for the event sample $\Scbetal$. Numbers in brackets indicate the ratio of confined to eruptive flares within the tested sample. Blue and red color indicates confined and eruptive flare type, respectively. X-class and M-class flares are shown as stars and circles, respectively.}
\label{fig:preflare_scatter_major}
\end{figure}

The detailed analysis of the preflare time profiles in Sect.~\href{sss:pre_long}{\ref{sss:pre_long}} suggests that one may use critical values to mark coronal conditions favorable for eruptive flaring ($\Efnp$\,=\,0.16, $\Hjn$\,=\,0.1, and $\Hjfn$\,=\,2.5\,$\times$\,$10^{-3}$; in close correspondence to those suggested in \cite{2021A&A...653A..69G} based on a small sample of ten flares). Aiming at the testing of the applicability of this particular set of threshold values, we inspect the distribution of the corresponding immediate preflare values for event sample $\Smajor$, and in particular their distribution with respect to different value regimes (quadrants; labeled "Q1" -- "Q4"), defined based on the thresholds $\Efnp$\,=\,0.16 and $\Hjfn$\,=\,2.5\,$\times$\,$10^{-3}$  (Fig.~\href{fig:preflare_scatter_major}{\ref{fig:preflare_scatter_major}}a). We find 22 events (21 eruptive) in Q1, no events in Q2, 25 events (11 eruptive) in Q3, and three eruptive events in Q4. Of the 21 events with $\Hjfn$ above the critical value (\ie, in Q1 and Q2) 21 (\ie, 95.5\%) were eruptive flares, and of the 24 events where $\Efnp$ exceeds the critical value (\ie, in Q1 and Q4) 24 (\ie, 96\%) were associated to a CME. We find a very similar situation when using the critical value $\Hjn$\,=\,0.1  (Fig.~\href{fig:preflare_scatter_major}{\ref{fig:preflare_scatter_major}}b), where from the 22 events above the critical value (\ie, in Q1 and Q2) 20 (\ie, 95.2\%) produced a CME. The flare type is predicted correctly in $\approx$\,70\% and 68\% of the events when the critical values of $\Hjn$ and $\Hjfn$ is used as an individual criterion, respectively (see sample $\Smajor$ in Table\,\href{table:preflare_scatter}{\ref{table:preflare_scatter}}). Only when requiring that $\Efnp$ simultaneously exceeds the critical value, the fraction of successfully predicted events is raised to $\approx$\,76\% of the events, and independent of the joint use with $\Hjn$ or $\Hjfn$. 

Our finding that the helicity-related relative proxies perform similarly successful in predicting the type of upcoming flaring clearly contrasts the statement by \cite{2023ApJ...945..102D} that $\Hjn$ is of lesser predictive power than $\Hjfn$. Therefore we need to discuss the significance of the thresholds suggested to indicate favorable conditions for eruptive flaring. One aspect to consider is the composition of the studied event sample. To do so, we assess the rate of successful prediction of the flare type for event samples composed differently. Besides the already analyzed sample $\Smajor$, we repeat our analysis also for a sample composed of smaller M-class flares only (up to GOES class M4.9; labeled "$\Sminor$" in Table\,\href{table:preflare_scatter}{\ref{table:preflare_scatter}}), a sample composed of X-class flares only (labeled "$\Sxonly$"), and a sample which covers all 232 flares (labeled "$\Sall$"). From Table\,\href{table:preflare_scatter}{\ref{table:preflare_scatter}} it appears that, when used as an individual measure, $\Hjn$ performs slightly better than $\Hjfn$ for event samples with a considerable fraction of X-class flares ($\gtrsim$\,44\%; see our event samples $\Smajor$, $\Scbetal$, and $\Sxonly$). For completeness we note here that the fraction of X-class flares in the event sample used in \cite{2023ApJ...945..102D} is $\approx$\,36\%, a bit lower as the fraction of X-class flares in our sample $\Ssea$ ($\approx$\,38\%) for which we find $\Hjn$ and $\Hjfn$ to preform equally successful. Therefore, we suspect that the reason why they attested a lower predictive ability to $\Hjn$ is rooted in the fact that their flare sample consisted of a larger number of smaller flares (they also included flares of GOES class M4 into their sample of major flares). Another aspect to consider is certainly the method used to perform the NLFF modeling (the CESE-MHD-NLFF method in their work). From comparative inspection of the values listed in Table~1 of \cite{2023ApJ...945..102D} it appears that while our estimates of $\tilde\phi$ are nearly 1:1 correlated, their estimates of $\Hj$ are larger by a factor of $\sim$7, which naturally results in larger estimates of $\Hjfn$ and thus a larger suggested critical value ($\Hjfn$\,=\,9\,$\times$\,$10^{-3}$). It is unclear, however, how many of the analyzed models in their work do actually qualify for helicity computation (hence do represent reliable data points in their statistics) since the authors omit to explicitly demonstrate the corresponding qualification of the underlying magnetic field models using a sufficiently sensitive metric to do so \citep[such as $\Edivprime$; for a dedicated work see][]{2022A&A...662A...3T}.

It has been long suggested in related research that a successful prediction of the CME-productivity of solar ARs necessarily involves a quantification of the restriction of the background (strapping) field \citep[\eg,][]{2012ApJ...748...77S}. With the hope to possibly improve the success rate of flare type prediction in that way, we therefore include information of the stabilizing effect of the strapping magnetic field in the preflare configurations. In particular, we inspect the critical height for torus instability ($\hcrit$) which is based on the rate at which the background field decreases with height. The latter is quantified by the decay index, $n$\,=\,$-\partial(\ln\vBh)/\partial(\ln z)$, where $\vBh$ is the horizontal component of the strapping field and $z$ is the height above the photosphere \citep{2006PhRvL..96y5002K}. For further analysis we extract the values of $\hcrit$ from the statistical analysis of \cite{2018ApJ...853..105B}, which covers 37 of the major flares of our sample $\Smajor$.

\begin{table}
\footnotesize
\setlength{\tabcolsep}{2.5pt}
\caption{Success rate of flare-type prediction for differently composed flare samples.} 
\label{table:preflare_scatter}
\centering
\begin{tabular}{lccccc}%
\hline\hline
Flare & No.~of & \multicolumn{2}{c}{Sample composition} & \multicolumn{2}{c}{Success rate (\%)} \\    
sample & events  & X-class & M-class & $\Hjn$ & $\Hjfn$\\
\hline
$\Sall$ &    231 & 22 (6/16) & 210 (120/90) & 57.1 & 58.4 \\ 
$\Sminor$ &    181 & $\times$ & 181 (111/70) & 53.6 & 55.8 \\ 
$\Smajor$ &     ~~50 & 22 (6/16) & 28 ~~(9/19) & 70.0 & 68.0 \\ 
$\Scbetal$ &    ~~37 & 19 (5/14) & 18 ~~(5/13) & 70.3 & 67.6 \\ 
$\Sxonly$ &    ~~22 & 22 (6/16) & $\times$ & 68.2 & 63.6\\ 
\hline
\end{tabular}
\tablefoot{
Event samples are composed as follows. $\Sall$: 231 flares of GOES class M1 or larger. S$_{M4-}$: smaller M-class flares (up to GOES class M4). $\Smajor$: major flares (GOES class M5 or larger). $\Sxonly$: X-class flares (GOES Xlass M1 or larger). Numbers in brackets indicate the ratio of confined to eruptive flares within the sample. Success rates are given in percent of the total number of flares within a sample. For flare-type prediction the pre-flare values of $\Hjn$\,=\,0.1 or $\Hjfn$\,=\,2.5\,$\times$\,$10^{-3}$ were used as an individual criterion.
}
\end{table}

The corresponding distribution of preflare values suggests a stronger segregation in terms of $\hcrit$ than of $\Hjn$ (see Fig.~\href{fig:preflare_scatter_major}{\ref{fig:preflare_scatter_major}}c), in support of the recently presented analysis of a small number of flares by \cite{2023A&A...volA..ppG}. More precisely, out of the 37 flares of sample $\Scbetal$, 28 are associated to $\hcrit$\,<\,40\,Mm, out of which 92.9\% were eruptive. For these 37 flares, the flare type is predicted correctly in 70.3\% when using the $\Hjn$\,$\ge$\,0.1 as condition for eruptivity and in 67.6\% when using $\Hjfn$\,=\,2.5\,$\times$\,$10^{-3}$ as individual criterion (see Table\,\href{table:preflare_scatter}{\ref{table:preflare_scatter}}). Using $\Efnp$\,$\ge$\,0.16 as an additional criterion, the flare type is predicted correctly in 78.4\% of the events. In comparison, requiring a preflare value of $\hcrit$\,<\,40\,Mm as an additional criterion to $\Hjn$\,$\ge$\,0.1 or $\Hjfn$\,=\,2.5\,$\times$\,$10^{-3}$ yields a drastically enhanced success rate of 94.6\% for the prediction of the flare type. In summary, (1) highest success rates regarding the prediction of the CME-association of upcoming flaring is the joint use of critical values of either $\Hjn$ or $\Hjfn$, in addition to that of $\hcrit$ as selection criteria, (2) the latter being the crucial quantity regarding flare type segregation. Finding (1) above is in agreement with the results presented in \cite{2022ApJ...926L..14L} who assessed the probability for a large flare to be associated with a CME based on a newly introduced parameter, $\alpha_{\rm FPIL}$/$\Fabs$, to quantify the relative importance of the magnetic nonpotentiality of an AR and the constraining effect of background magnetic field (see their Sect.~2 for details). They defined $\alpha_{\rm FPIL}$ as being the mean twist parameter $\alpha$\,=\,$\mu$\,$\sum$$J_z$\,$B_z$/$\sum$$B_z^2$ employed only for the area covered by the flare polarity inversion line (FPIL). They found the ratio to exceed critical values in about 90\% of eruptive flares (43 events in their study).

A related topic which we might touch upon is the relation of the helicity (proxies) to the overall magnetic setting of the host AR. Dedicated analyses suggested that $\hcrit$ can be roughly approximated by half of the distance between the flux-weighted centers of opposite polarity ($\dpc$) in the host AR \citep[\eg,][]{2018ApJ...853..105B,2022A&A...665A..37J}. Taking this rough approximation into consideration, Fig.~\href{fig:preflare_scatter_major}{\ref{fig:preflare_scatter_major}}c implies that $\Hjn$ is smaller for more extended ARs (with $\Hjn$\,$\lesssim$\,0.05 for ARs with $\dpc$\,$\gtrsim$\,60\,Mm; not shown explicitly). We show the flare distance ($\dfc$; defined as the distance between the flare site and the flux-weighted center of the AR), normalized with respect to the geometrical extent of the active-region magnetic dipole (accomplished through division by $\dpc$/2). A value $\dfcprime$\,$<$\,1 refers to a location underneath the magnetic field connecting the flux-weighted centers of opposite polarity (called the confining dipole field, hereafter). As already noted by \cite{2018ApJ...853..105B}, confined flares are found to reside preferentially at such locations in extended ARs ($\dpc$\,$\gtrsim$\,60\,Mm). From our joint analysis with $\Hjn$ we are now able to show that the corresponding host AR exhibits values of $\Hjn$ below the critical value (see lower left quadrant in Fig.~\href{fig:preflare_scatter_major}{\ref{fig:preflare_scatter_major}}d). In contrast, from underneath the confining dipole field in compact ARs ($\dpc$\,$\lesssim$\,60\,Mm) almost exclusively eruptive flares originate, with their preflare corona being characterized by values of $\Hjn$ exceeding the critical threshold (see lower right quadrant). The largest fraction of major events in such a representation, however, originates from locations in the periphery of the host AR ($\dfcprime$\,$>$\,1), and exhibits a broad distribution of preflare values of $\Hjn$.

\begin{figure}
\centering
\includegraphics[width=\columnwidth]{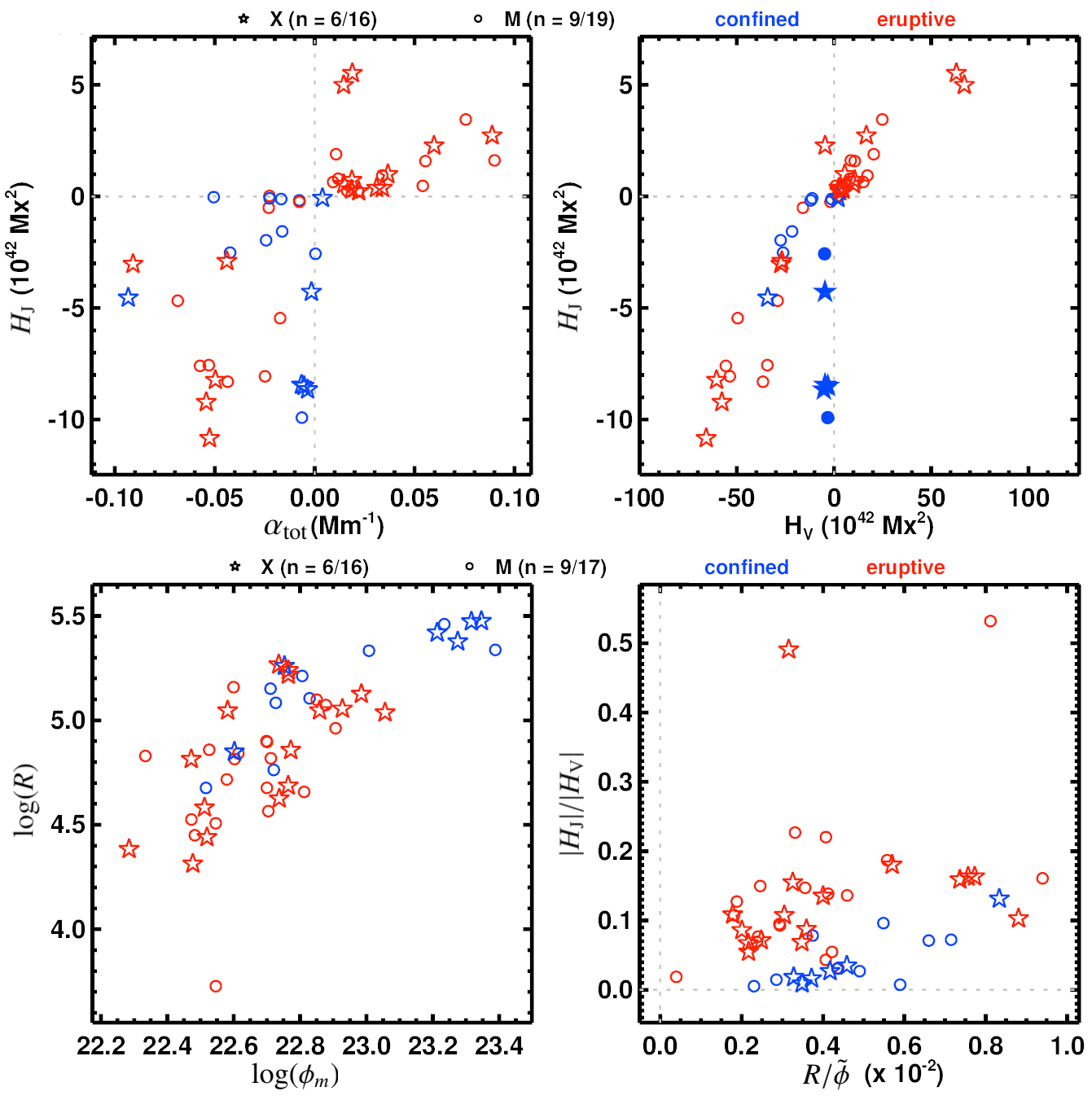}
\caption{Preflare values of selected photospheric and coronal measures for 48 flares of sample $\Smajor$. (a) Helicity of the current-carrying field vs.\ mean characteristic twist parameter, $\alpha_{\rm total}$ and (b) near-PIL magnetic flux measure, $\log(R)$, value vs.\ total magnetic flux, $\log(\Fabs)$ for the flares of sample $\Smajor$. (c) Flux-$R$ vs.\ $\log$$\Fabs$, and (d) helicity ratio, $\Hjn$, vs.\ flux-normalized flux-$R$ ($\Rprime$) (events with $\log(R)$\,$<$1 are excluded). Numbers in brackets indicate the ratio of confined to eruptive flares within the tested sample. Blue and red color indicates confined and eruptive flare type, respectively. X-class and M-class flares are shown as stars and circles, respectively.}
\label{fig:jsoc_scatter}
\end{figure}

The quantity $\alpha$ introduced above is often used to parameterize the surface magnetic field in ARs and often serves as a measure for the coronal helicity \citep[\eg,][]{1995ApJ...440L.109P,2008ApJ...673..532T}. HMI SHARP data, the basis for our employed NLFF models, contain a number of regularly provided keywords \citep[cf.\ Table~3 of][]{2014SoPh..289.3549B}, including $\alphat$. Therefore it is straightforward to compare this proxy to our FV helicity estimates. For our sample $\Smajor$ we find $\alphat$ being moderately indicative for the amount of helicity present (see Fig.~\href{fig:jsoc_scatter}{\ref{fig:jsoc_scatter}}a, where we use $\Hj$ as a reference). We find Spearman rank correlations of $\Csr$\,=\,0.67 ($P$\,$<$\,0.01)\footnote{The probability ($P$ in the interval [0,1]) indicates whether a correlation could have occurred by random chance, with smaller values indicating a significant correlation. The confidence level or statistical significance of a correlation can then be expressed as (1\,$-$\,$P$)\,$\times$\,100\%, and a significance of $\ge$\,95\% is conventionally considered to be statistically significant. Only when explicitly noted statistical significance is not guaranteed.
}
and 0.78 with respect to the distributions of $\Hv$ and $\Hj$, respectively.\footnote{We use the following guide to categorize the strength of a correlation: $\Csr$\,$\le$\,0.59 -- weak, $\Csr$\,$\in$\,[0.6, 0.79] -- moderate, and $\Csr$\,$\in$\,[0.8, 1.0] -- strong.
}
Furthermore, we find that $\alphat$ shares the sign of $\Hv$ in 46 of the cases (\ie, 92\%) and shares the sign of $\Hj$ in 47 of the cases (\ie, 94\%). Considering all flares of GOES class M1 or larger (sample $\Sall$), these fractions reduce to 87.9\% and 91.4\%, respectively. In comparison, \cite{2014ApJ...785...13L} estimated that the sign of the helicity and the force-free parameter in 28 newly emerging ARs (based on the assessment of the photospheric helicity flux and using a weighted mean estimate of $\alphat$) agrees in $\gtrsim$\,89$\pm$11\% of events. Without going into detail regarding the challenges of employing a meaningful measure of $\alphat$ (for this we refer to the a dedicated discussion in Sect.\ 4.1 of \cite{2014ApJ...785...13L}) we do suggest (1) that $\alphat$ is a slightly better proxy of the helicity of the current-carrying field, $\Hj$, and (2) that one might fail to correctly guess the sign of the coronal helicity prior to large flares for a small fraction of events (at least 10\%). In this context it is to be noted, however, that though $\Hv$ and $\Hj$ are strongly correlated they do not necessarily share the same sign (see Fig.~\href{fig:jsoc_scatter}{\ref{fig:jsoc_scatter}}b). In cases where the overall handedness of the host AR magnetic field is the same as that of the nonpotential (current-carrying) structure the signs are likely the same. We find $\Hv$ and $\Hj$ to share the same sign in (97.1\%) 94\% of the (eruptive) major flares. Considering all flares of GOES class M1 or larger (sample $\Sall$), these fractions reduce to (90.6\%) 89.7\% of (eruptive) events. This supports that for the majority of flares the preflare helicity is linked to a (single) prominent current-carrying magnetic structure which determines the overall structural complexity of the host AR.

Another photospheric measure that has been suggest to be highly indicative of the flare potential of solar ARs is related to the emergence of magnetic flux exhibiting strong-field, high-gradient polarity inversion lines \citep[flux-$R$;][]{2007ApJ...655L.117S}. Similar to the findings in that pioneering study, we find major flaring associated with a large range of total unsigned fluxes of the host AR, along with values of $\log(R)$\,$\gtrsim$\,3.5 (see Fig.~\href{fig:jsoc_scatter}{\ref{fig:jsoc_scatter}}c). From this representation it is clear that the flux-$R$ does not distinguish the major flare type (confined/eruptive flares are shown in blue/red color), nor is it related to the flare size (M- and X-class flares are represented by circles and stars, respectively). A recent in-depth study by \cite{2024A&A...683A..87M} showed that the flux-$R$ correlates well with a similar yet helicity-related measure, $R_H$, but that the latter is more indicative of eruptive flaring, in line with the fact that helicity is more indicative than magnetic flux in general \citep[][]{2022A&A...662A...3T,2022A&A...662A...6L}. Assessing the relation between the flux-$R$ and our volume-based helicity proxy, $\Hjn$, we employ the flux-normalized measure $\Rprime$ and find it little indicative of the type of upcoming flaring (see Fig.~\href{fig:jsoc_scatter}{\ref{fig:jsoc_scatter}}d). Also, we find it to be essentially uncorrelated with $\Hjn$ ($\Csr$\,=\,0.25; along with a significance of only $\sim$92\%). 

Finally, to bridge to the comparative analysis of the long-term preflare time evolution in Sect.~\href{sss:dtw}{\ref{sss:dtw}}, we look at the Pearson rank correlations of the immediate preflare values (see Table\,\href{table:preflare_corr}{\ref{table:preflare_corr}}). Drawn from our sample $\Smajor$, similarities with earlier studies include strongest correlations between the unsigned photospheric flux ($\Fabs$) and the total energy ($\Csr$\,=\,0.94), between the unsigned values of $\Hvabs$ and $\Hjabs$ ($\Csr$\,=\,0.89), as well as between the free energy ($\Ef$) and the unsigned total helicity and as well as the unsigned helicity of the current-carrying field ($\Csr$\,$\gtrsim$\,0.9). Importantly, this strong correlation between the immediate preflare values of $\Ef$ and $\Hjabs$ is not reflected in a correspondingly low cost during the long-term (24-hour) preflare time evolution ($\mCcum$($\Ef$,$\Hjabs$)\,=\,2.57$\pm$0.13 in Table\,\href{table:preflare_cost}{\ref{table:preflare_cost}}), highlighting the limitations of using single-valued correlation coefficients for the quantification of the nearness of two data samples.

\begin{table}
\scriptsize
\setlength{\tabcolsep}{2.5pt}
\caption{Spearman rank correlation of photospheric and coronal quantities prior to major flares.}
\label{table:preflare_corr}
\centering
\begin{tabularx}{\columnwidth} { 
  | >{\raggedright\arraybackslash}X 
  | >{\centering\arraybackslash}X 
  | >{\centering\arraybackslash}X 
  | >{\centering\arraybackslash}X 
  | >{\centering\arraybackslash}X 
  | >{\raggedleft\arraybackslash}X | 
  }
\hline
~ & $\Fabs$  & $\Et$ & $\Ef$ & $\Hvabs$ & $\Hjabs$\\ 
\hline
$\Fabs$ & 1.00 & \scol 0.94 & \mcol 0.66 & \mcol 0.70 & \wcol 0.52 \\
\hline
$\Et$ & ~ & 1.00 & \mcol 0.75 & \scol 0.81 & \mcol 0.62 \\
\hline
$\Ef$ & ~ & ~ & 1.00 & \scol 0.90 & \scol 0.94 \\
\hline
$\Hvabs$ & ~ & ~ & ~ & 1.00 & \scol 0.89 \\
\hline
$\Hjabs$ & ~ & ~ & ~ & ~ & 1.00 \\
\hline
\end{tabularx}
~\\~\\
\begin{tabularx}{\columnwidth} { 
  | >{\raggedright\arraybackslash}X 
  | >{\centering\arraybackslash}X 
  | >{\centering\arraybackslash}X 
  | >{\raggedleft\arraybackslash}X | 
  }
\hline
~ & $\Efnp$ & $\Hjn$ & $\Hjfn$\\ 
\hline
$\Efnp$ & 1.00 & \scol 0.88 & \scol 0.90 \\
\hline
$\Hjn$ & ~ & 1.00 & \scol 0.82\\
\hline
$\Hjfn$ & ~ & ~ & 1.00 \\ 
\hline
\end{tabularx}
\tablefoot{
Mean values are computed from the pre-flare values of $\Csr$ of all flares of sample $\Smajor$. The cell colors are indicating the strength of the correlation and refer to a correlation of $\Csr$\,$\geq$\,0.8 (dark red) as "strong", 0.8\,$>$\,$\Csr$\,$\geq$\,0.6 (orange) as "moderate", and $\Csr$\,$<$\,0.6 (light red) as "weak".
}
\end{table}

\subsection{Flare-related changes}
\label{ss:changes}

\begin{figure}
\centering
\includegraphics[width=\columnwidth]{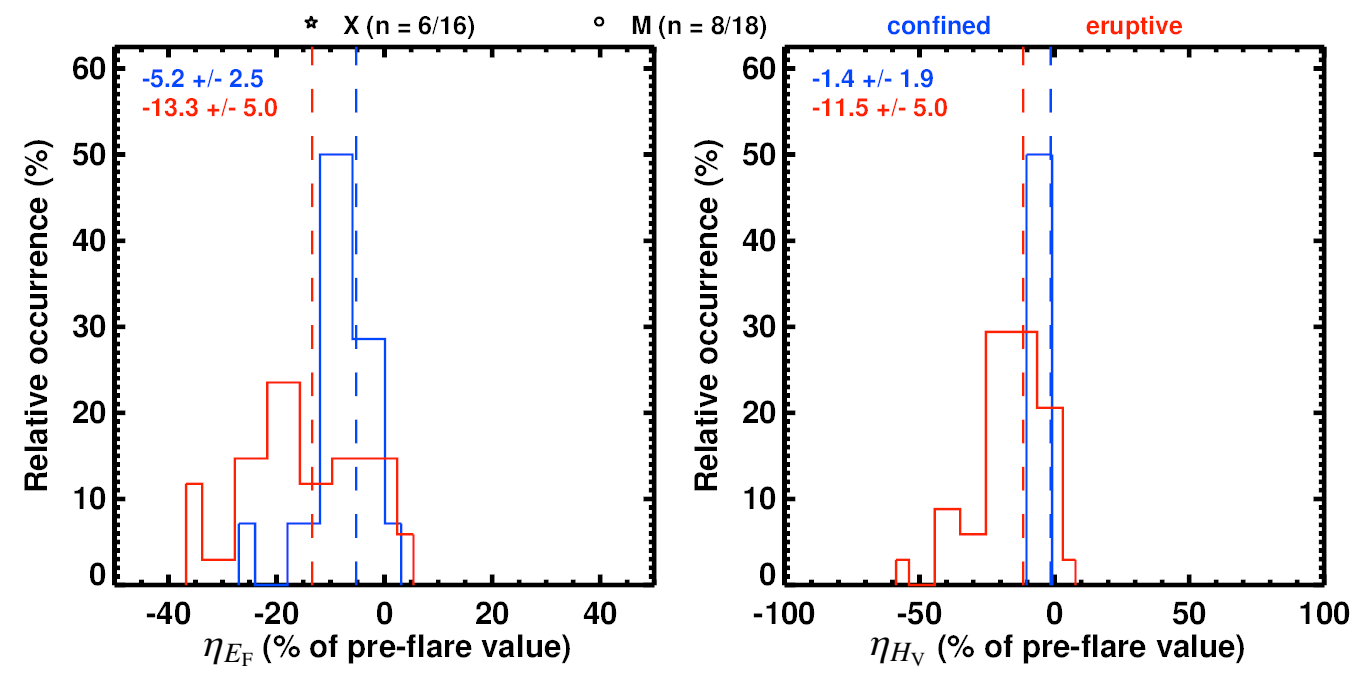}
\caption{Histograms of mean relative flare-related changes to the (a) free magnetic energy ($\eta_{\Ef}$) and (b) total helicity ($\eta_{\Hv}$) for the sample $\Smajor$. Numbers in brackets indicate the ratio of confined to eruptive flares within the tested sample. Flares for which a helicity reversal is detected are excluded. Blue and red color indicates confined and eruptive flare type, respectively.}
\label{fig:histo_changes}
\end{figure}

Based the analysis of the flare-related changes in our sample $\Smajor$ with respect to the preflare budget supports a flare-type related distinction of flare-related changes. We evaluate the flare-related changes with respect to the preflare content in the form $\eta_{\Ef}$\,=\,$\langle\Efpost$\,$-$\,$\Efpre\rangle$/$\langle\Efpre\rangle$ and $\eta_{\Hv}$\,=\,$\langle\Hvpost$\,$-$\,$\Hvpre\rangle$/$\langle\Hvpre\rangle$, where angular brackets denote one-hour time averages. The preevent maps cover one hour prior to $t$\,=\,($t_0$\,$-$\,10\,min) and the postevent maps cover one hour after the nominal GOES end time. We note here that we exclude flares from analysis for which we detect a reversal in the sign of $\Hv$ or $\Hj$ since then it is difficult to define a reference preflare levels for the flare-related changes in a meaningful way. For our sample $\Sall$ this concerns 11 flares, out of which two were major flares (SOL2012-05-10T04:11M5.7 and SOL2013-04-11T06:55M6.5). We find mean values of $|\eta_{\Ef}|$\,=\,13.3$\pm$5.0\% and $|\eta_H|$\,=\,11.5$\pm$5.0\% for eruptive flares, compared to $|\eta_{\Ef}|$\,=\,5.2$\pm$2.5\% and $|\eta_H|$\,=\,1.4$\pm$1.9\% for confined flares (see Fig.~\href{fig:histo_changes}{\ref{fig:histo_changes}}a and Fig.~\href{fig:histo_changes}{\ref{fig:histo_changes}}b, respectively).
In words, (1) a considerable amount of both free magnetic energy and helicity ($\gtrsim$\,10\%) is liberated during major eruptive flares, and (2) the overall helicity budget is more or less conserved during confined events ($|\eta_{\Hv}|$\,$\approx$\,1\%). For the helicity budget of the current-carrying field we find, besides the expected pronounced eruptive-flare related modification ($|\eta_{\Hj}|$\,=\,18.2$\pm$6.5\%, also a considerable amount that is being processed during confined flares (9.0$\pm$4.1\%). We note here that this strong change to the budget of $\Hj$ is mainly related to X-class flaring (for discussion see Sect.~\href{ss:post}{\ref{ss:post}}).

Our findings are in support of the recent analysis of \cite{2023RAA....23i5025W} who found the flare-related changes due to eruptive flares ($\eta_E$\,$\sim$\,$\eta_H$\,$\sim$\,15\%) significantly larger when compared to that for confined flares ($\eta_E$\,$\sim$\,4\% and $\eta_H$\,$\sim$\,1\%), based on linear fittings to the values over four data points prior to/after the flare start/end times. The authors left out the chance, however, to provide quantitative details of their model results which would demonstrate whether or not they do numerically qualify for subsequent helicity computation. The authors argue for the validity of their modeling with presenting small numbers of the fractional flux metric, as introduced in \cite{2000ApJ...540.1150W}, which is known to depend on the spatial sampling of the data \citep[see dedicated work by][]{2020ApJ...900..136G}. \cite{2022A&A...662A...3T} extensively discussed the nonability of the  fractional flux metric in testing the solenoidal qualification of a NLFF model for subsequent helicity computation and the need to use alternative measures \citep[such as $\Edivprime$; for a dedicated work see][]{2022A&A...662A...3T}. Therefore, it remains unclear how many of the analyzed helicity values in \cite{2023RAA....23i5025W} do actually represent valid data points. Further, \cite{2023RAA....23i5025W} use original-resolution HMI data as input to magnetic field modeling as an argument for enhanced realism and quality of NLFF modeling. The latter, again, cannot be reconciled since it is known that the inclusion of "more" real information (via a finer spatial sampling) does not guarantee a more internally consistent solution by itself \citep[for a comparative study see][]{2009ApJ...696.1780D}. Instead, for the optimization method used in the study of \cite{2023RAA....23i5025W}, it has actually been demonstrated by \cite{2022A&A...662A...3T} that an improved (finer) spatial sampling lowers the final NLFF model quality. More precisely, they showed that NLFF quality tends to be higher for larger pixel sizes (\ie, binned data) when using $\thetaj$ and $\Edivprime$ as measures. In addition they examined that for HMI data binned to a plate scale of $\sim$\,0.72\,Mm (as used in this study) the changes to the deduced magnetic energies and helicities are small compared to other possible sources of uncertainty (see their Sect.~5).

\begin{figure}
\centering
\includegraphics[width=\columnwidth]{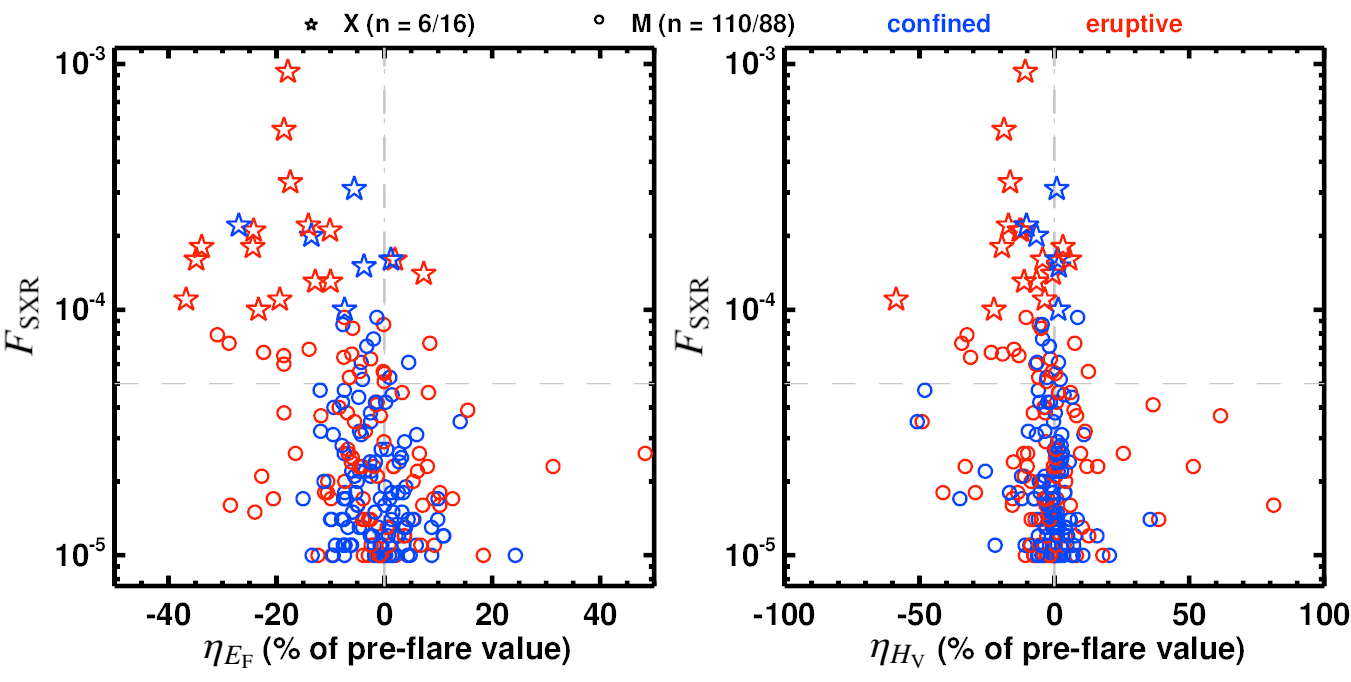}
\caption{GOES 1--8\,\AA peak flux ($F_{\rm SXR}$) as function of the mean flare-related changes for 220 flares of sample $\Sall$. The horizontal dashed lines indicate a SXR flux of 5\,$\times$\,$10^{-5}$\,W\,m$^{-2}$, i.e., GOES class M5. Numbers in brackets indicate the ratio of confined to eruptive flares within the tested sample. Flares for which a helicity reversal is detected were excluded. Blue and red color indicates confined and eruptive flare type, respectively. X-class and M-class flares are shown as stars and circles, respectively.}
\label{fig:scatter_changes}
\end{figure}

Next, we inspect the flare-related changes for 220 large flares of our sample $\Sall$ (see Fig.~\href{fig:scatter_changes}{\ref{fig:scatter_changes}}). For the subset of 104 eruptive flares, we find flare-related reductions to $\Ef$, $\Hv$, and $\Hj$ (not necessarily simultaneously) in $\sim$\,72.1\%, $\sim$\,66.3\%, and $\sim$\,62.1\% of the events, respectively. These percentages increase to $\sim$\,91.2\%, $\sim$\,85.3\%, and $\sim$\,91.2\%, respectively, when only major flares (of GOES class M5 or above) are considered (see plot symbols above the horizontal dashed lines in Fig.~\href{fig:scatter_changes}{\ref{fig:scatter_changes}}). For our sample $\Sall$, we find a simultaneous decrease in $\Ef$ and $\Hjabs$ in $\approx$\,62\% of the eruptive flares, compared to $\approx$\,51\% for confined flares. These fractions increase to $\approx$\,88\% and $\approx$\,71\% when considering only major (of GOES class M5 or above) eruptive and confined flares, respectively.

Notably, there are also a considerable number (75) of events for which we find an increase of $\Ef$ during flares with the number increasing with decreasing peak flare flux (see Fig.~\href{fig:scatter_changes}{\ref{fig:scatter_changes}}a). In 69 of these events, the flare-related changes are either below the "background" variation of $\Ef$ (d$\Ef$/d$t$\,$\lesssim$\,1\,$\times$\,$10^{31}$\,$\erg$\,h$^{-1}$; cf.\, Fig.~\href{fig:grad1h_sea_major}{\ref{fig:grad1h_sea_major}} in Sect.~\href{sss:grad}{\ref{sss:grad}}), or d$\Ef$/d$t$\,$<$\,10\%, that is, below the characteristic uncertainty of $\Ef$. The latter has been found in different studies targeted on the application of different FV helicity methods \citep[see][and Fig.~2 therein]{2019ApJ...887...64T}, or the effect of the spatial sampling of the input data \citep[see][and Fig.~7 therein]{2022A&A...662A...3T}. For the remaining seven events (GOES classes M1 or M2) the flare-related changes are exceeding the "background variations" as well as are larger than 10\% of the mean preflare level. Upon visual inspection of the respective time profiles they are found in association with an increasing total and potential energy during the flares. 

\subsection{Postflare evolution}\label{ss:post}

\begin{figure*}
\centering
\includegraphics[width=0.75\textwidth]{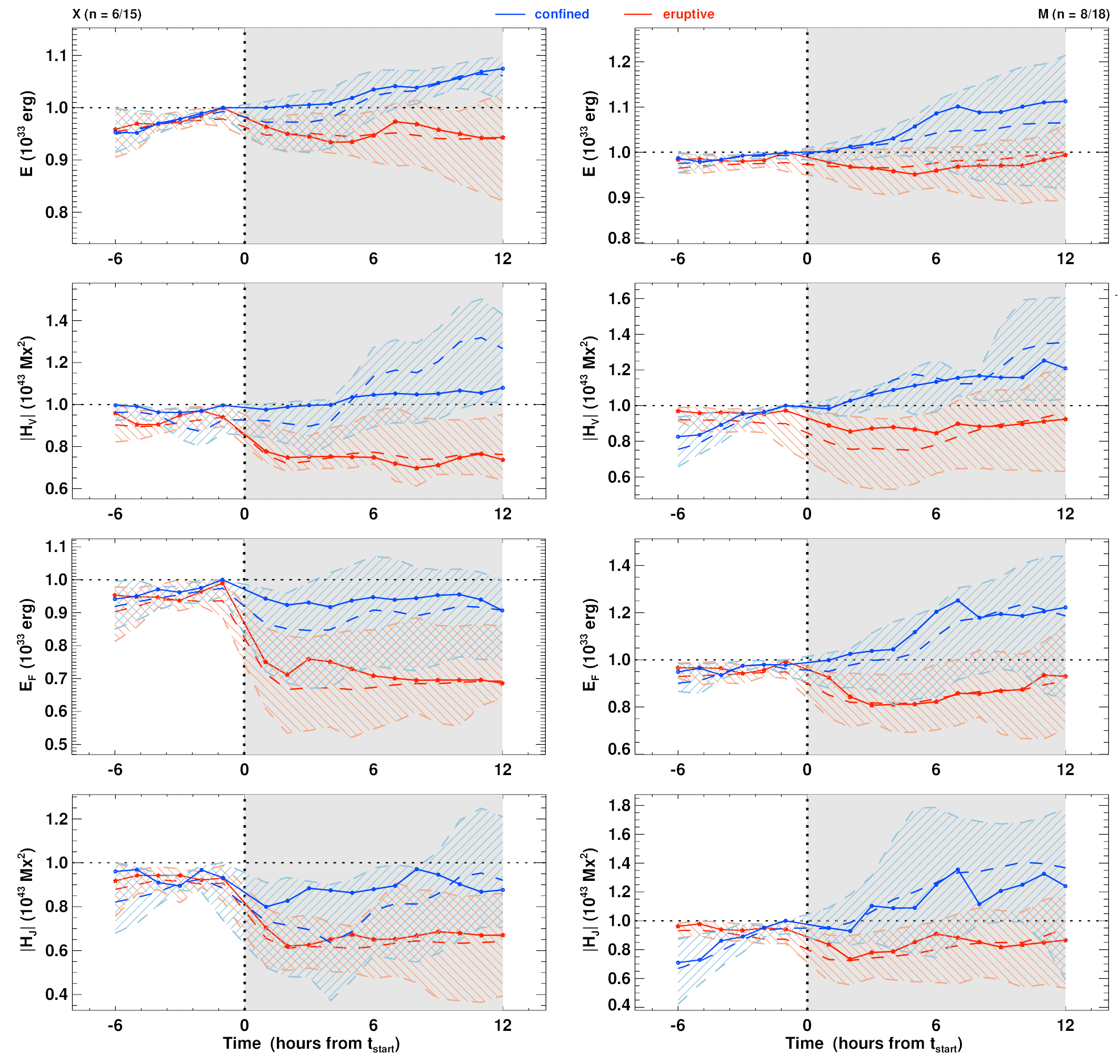}
\caption{Average time evolution of selected quantities computed from the normalized (with respect to the respective preflare maximum values) superposed epoch data of event sample $\Sxonly$ (only X-class flares; left column) and for large M-class flares (GOES class M5 to M9; right column), separately for the subsets of confined (blue) and eruptive (red) flares. Numbers in brackets indicate the ratio of confined to eruptive flares within the tested sample. Covered is the time period from $-$6~hours to $+$12~hours around the flare start time (black dotted vertical line). The major-flare-less postflare time period of 12~hours used for selection of qualified events is marked by the gray-shaded area. From top to bottom in the each column the mean and median values of the total magnetic energy ($\Et$), free energy ($\Ef$), total unsigned magnetic helicity ($\Hvabs$), and unsigned helicity of the current-carrying field ($\Hjabs$) are shown. Dashed lines represent mean values. Solid lines connecting symbols represent median values. The shaded regimes are bound by the upper and lower quartiles of the respective distributions. Horizontal lines indicate a characteristic preflare level.}
\label{fig:mnsea_post}
\end{figure*}

To study the effects of flares on the magnetic energy and helicity budgets of the host ARs, we applied SEA to the interval $-$6~hours to $+$12~hours around the flare start time (Fig.~\href{fig:mnsea_post}{\ref{fig:mnsea_post}}). We did not require any major flare to occur in the postflare interval to reduce the effect of subsequent flaring onto the synchronized time profiles and to avoid duplicate epochs in the superposition. We applied SEA to the normalized time profiles to allow for a straightforward interpretation of typical flare-related changes with respect to characteristic preflare levels. For the normalization,  for each parameter we used the maximum value during the preflare time interval. We analyzed two samples of flares separately, once including only X-flares (6 confined and 15 eruptive events) and once including only large M-class flares (GOES classes M5 to M9; 8 confined and 18 eruptive flares). In addition, we considered both mean and median values in order to account for the different fractions of confined and eruptive flares in the two tested flare samples. This allowed us to differentiate between both different flare types and sizes (magnitudes) with respect to the particular time needed for replenishment. 

We find that the eruptive-flare related changes require $\sim$2 -- 3~hours to take full effect (both, for X-class as well as large M-class flares). Furthermore, consistent with the findings discussed in Sect.~\href{ss:changes}{\ref{ss:changes}}, we find that $\Hv$ is approximately conserved during magnetic reconnection (see blue curves in Fig.~\href{fig:mnsea_post}{\ref{fig:mnsea_post}}c and \href{fig:mnsea_post}{\ref{fig:mnsea_post}}d). This is also approximately true for $\Hjabs$ in the absence of X-class flaring (see blue curves in Fig.~\href{fig:mnsea_post}{\ref{fig:mnsea_post}}f). \cite{2023ApJ...942...27L} estimated that, on average, eruptive X-flares remove about 10\% of free energy available in the host AR and roughly 12~hours pass until the coronal budget returns to near-preflare values. A corresponding behavior in $\Hv$ was suggested to comply with the idea that the solar interior acts as a reservoir to resupply coronal helicity as it is removed by CMEs \citep[][]{2000ApJ...545.1089L}. Based on our refined analysis (considering only X-flares with a postflare time window of 12~hours free of other major flare occurrence) we find characteristic changes related to eruptive X-class flaring of $\lesssim$15\% and find a return of the budgets of $\Et$ and $\Hvabs$ to near-preflare levels to require at least $\sim$12~hours (see Fig.~\href{fig:mnsea_post}{\ref{fig:mnsea_post}}a and \href{fig:mnsea_post}{\ref{fig:mnsea_post}}c, respectively). A similar behavior is found for the postflare time evolutions of $\Ef$ and $\Hjabs$ (see Fig.~\href{fig:mnsea_post}{\ref{fig:mnsea_post}}e and \href{fig:mnsea_post}{\ref{fig:mnsea_post}}g, respectively), while the modifications to the budgets due to the occurrence of large eruptive flaring is more pronounced ($\approx$\,20--30\%).

In comparison, large eruptive M-class flares (GOES classes M5 to M9) affect the coronal budgets less strongly (see right column of Fig.~\href{fig:mnsea_post}{\ref{fig:mnsea_post}}), with lower mean reductions to the budgets of $\Et$ and $\Hvabs$ ($\lesssim$10\%) as well as of $\Ef$ and $\Hjabs$ ($\lesssim$\,20\%). The replenishment of the budgets is accomplished within roughly 12~hours, for $\Et$ and $\Hvabs$ possibly even within $\sim$6~hours. Notably, the median values of $\Ef$ indicate a slight increase during large confined M-class flares (see solid curve in Fig.~\href{fig:mnsea_post}{\ref{fig:mnsea_post}}f), while the mean values (dashed curve) indicate a slight decrease (consistent with the expectation that free energy is released during flares). The apparently inconsistent trend of the flare-related median values of $\Ef$ are rooted in the stronger effect of increases of $\Et$ (and hence apparent increases of $\Ef$) in two (SOL2012-07-04T09:47M5.3 and SOL2012-07-05T11:39M6.1) out of the eight 
analyzed confined flares. The (free) energy increases in those two flares were rooted in flux emergence in the host AR (NOAA~11515; see, \eg, \cite{2019ApJ...881..151L}). For completeness we note that smaller eruptive flares (GOES classes M1 to M4; not shown explicitly) have only little impact on the overall coronal budgets, with the replenishment being accomplished in practically no time. The validity of this finding is limited, however, as changes of the free energy and helicity budget associated to small M-class flares are often found in the range of the overall uncertainty of the underlying quantities (tested in, \eg, \cite{2019ApJ...887...64T} and \cite{2022A&A...662A...3T}) and/or are found smaller than their characteristic variations in the absence of major flaring (see Sect.~\href{sss:grad}{\ref{sss:grad}} for a detailed analysis).

\section{Extended discussion}
\label{s:discussion}

Our high-quality coronal magnetic field modeling has allowed us to cover the coronal evolution around 231 solar flares and investigate particular aspects, including (1) the long-term preflare time evolution, (2) the immediate preflare conditions, (3) flare-related changes, and (4) the postflare evolution of the coronal conditions. To address (1), (4), and partially (3), we synthesized information from selected major flare events using superposed epoch analysis \citep[SEA;][]{1913RSPTA.212...75C}. More specifically, the time series of selected quantities within predefined time windows around the flare start time were extracted and superposed after synchronization with respect to it. To avoid duplicate signals and misinterpretation of flare-related impacts we additionally required no major flare to happen within the studied time intervals. The resulting higher statistical reliability serves as an improvement to the pioneering study of \cite{2023ApJ...942...27L}. For the analysis of the time evolution of coronal quantities with respect to each other we apply dynamic time warping \citep[DTW; \eg,][]{2001SIAM...1K} to the superposed epoch data. To address (2), physical quantities were computed from the preflare NLFF models and analyzed, also compared to proxies for the flare ability of ARs often used in literature such as the flux-$R$ measure and overall AR twist, as well as the critical height for torus instability. This allowed us to study measures of the local stability in context with the global nonpotentiality of solar ARs, the latter approximated via the FV relative helicity. For addressing (3), mean values were computed from one-hour preflare and postflare time intervals and subsequently compared. 

In the following, we list our most notable results regarding the topical distinction defined above and discuss them with respect to literature. The different flare samples used to address the individual aspects are summarized in Table~\href{table_preflare_corr}{\ref{table:preflare_corr}}.

\subsection{Long-term preflare evolution}

    Over longer timescales -- during the 48~hours preceding major flares -- the overall time evolution of the total energy and helicity is not distinctly different prior to confined or eruptive flaring (see Fig.~\href{fig:msea_major}{\ref{fig:msea_major}}b and \href{fig:msea_major}{\ref{fig:msea_major}}d, respectively), in support of an earlier study by \cite{2023ApJ...942...27L}. Confined flares, however, originate from ARs with larger unsigned photospheric fluxes (Fig.~\href{fig:msea_major}{\ref{fig:msea_major}}a), and thus are associated to larger total energy and helicity budgets.
    
    The unsigned change rates of the coronal energy and helicity budgets prior to eruptive and confined flaring exhibit similar magnitudes, yet are somewhat larger prior to confined flaring (Fig.~\href{fig:grad1h_sea_major}{\ref{fig:grad1h_sea_major}}). On overall, the coronal change rates are about 100 times larger than the respective contributions from photospheric fluxes estimated in earlier works \citep[e.g.,][]{2014ApJ...785...13L,2020ApJ...904....6P}. Noteworthy, the change rates do not support a scenario of stagnation of the coronal helicity content prior to major flaring.  
    
    During the 24~hours leading to a major flare, the unsigned magnetic flux and the total magnetic energy evolve most similar with respect to each other, irrespective of the type of upcoming flaring (Table~\href{tab:preflare_cost}{\ref{table:preflare_cost}}).
    In principle, this supports that flux emergence is the most prominent driver of the coronal energy storage \citep[see also, \eg,][]{2012ApJ...748...77S}. We also find that there are dependencies on the type of upcoming flaring. For instance, prior to eruptive flares, $\Hjabs$ evolves more similar with respect to $\Fabs$ than prior to confined flares, while the opposite is true for the relative time evolution of $\Ef$ and $\Fabs$. A corresponding interpretation is challenging and requires more detailed and dedicated studies that aim to assess the contribution of photospheric drivers such as emerging or shearing motions to the accumulation of free energy and helicity over time in the corona above.
    
    Coronal conditions prior to eruptive and confined flaring are distinctly different regarding the overall levels of specific eruptivity proxies (free-to-potential energy ratio, helicity ratio, and flux-normalized helicity of the current-carrying field; see right column of Fig.~\href{fig:msea_major}{\ref{fig:msea_major}}), yet do their distributions exhibit a significant overlap. Thus, deduced "critical" values indicative of upcoming eruptive flaring are not to be thought of as universal. In any case, CME-favorable conditions do not develop just shortly before eruptive flaring is observed. Instead, they persist over extended time periods prior to eruptive flaring, possibly up to several days. Therefore, the exceeding of suggested critical values is not an indicator for near-in-time upcoming eruptive flaring.

\subsection{Immediate preflare conditions}

    We find favorable conditions for large flaring as $\Hv$\,$\gtrsim$\,1\,$\times$\,$10^{42}$\,$\mxmx$ (Fig.~\href{fig:preflare_scatter}{\ref{fig:preflare_scatter}}a) and $\Ef$\,$\gtrsim$\,2\,$\times$\,$10^{31}$\,$\erg$ (Fig.~\href{fig:preflare_scatter}{\ref{fig:preflare_scatter}}c). Our analysis of a statistically meaningful number of large flares 231 flares of GOES class M1 or larger) therefore supports earlier results based on smaller event samples \citep[\eg,][]{2012ApJ...759L...4T}. For comparison, based on the analysis of photospheric helicity and energy fluxes in newly emerging ARs, \cite{2022A&A...662A...6L} suggested thresholds for an AR to become CME-productive as $\Hv$\,$\gtrsim$\,9\,$\times$\,$10^{41}$\,$\mxmx$ and $\Et$\,$\gtrsim$\,2\,$\times$\,$10^{32}$\,$\erg$.

    There might exist an upper limit for the nonpotentiality a solar AR can host (Fig.~\href{fig:preflare_scatter}{\ref{fig:preflare_scatter}}a,b). More observations are needed to substantiate this observed aspect since it stems from the data associated to a single solar AR (NOAA~12192) which until today is outstanding regarding the extent it covered on the solar surface \citep[for a dedicated study see, \eg,][]{2012A&A...543A..49C} and the magnetic flux it hosted \citep[for a comparative study see, \eg,][]{2015ApJ...804L..28S}. 
    
    For the majority of major flares ($\approx$94\%; and for eruptive events even $\gtrsim$\,97\%), the preflare helicity is linked to nonpotential magnetic field which determines the overall structural complexity of the host AR.
    
    We cannot reconcile the suggestion of an earlier work in that $\Hjn$ is of lesser predictive power than $\Hjfn$ \citep{2023ApJ...945..102D} regarding the flare type. Instead it is clearly contrasted by our in-depth analysis that both have a similar ability to indicate the type of upcoming flaring, and when considering only largest (X-class) flares, $\Hjn$ is actually outperforming $\Hjfn$ (Table\,\href{table:preflare_scatter}{\ref{table:preflare_scatter}} and see also Fig.~\href{fig:preflare_scatter_major}{\ref{fig:preflare_scatter_major}}a,b).\\
    To substantiate our findings, we explicitly demonstrate how the obtained success rate depends on (i) the event sample studied (and in particular the fraction of X-class flares contained), (ii) the critical values used which in turn is certainly dependent on the method used to model the 3D magnetic field \citep[see, e.g., supplementary material of][]{2023NatAs...7.1171J}, and (iii) the qualification of the employed magnetic field modeling for subsequent helicity computation, based on which we attest our findings a higher reliability than that of earlier studies (see detailed discussion in Sect.~\href{ss:pre}{\ref{ss:pre}}).
    
    We solidify the suggestion of a pioneering study of \cite{2023A&A...volA..ppG} -- who analyzed a limited set of ten flares -- that local measures of coronal stability (e.g., $\hcrit$) are much more segregate in terms of flare type than the helicity-related measures of the global nonpotentiality ("global" in the sense of covering the active-region corona; see Fig.~\href{fig:preflare_scatter_major}{\ref{fig:preflare_scatter_major}}c). Most importantly, a combination of these proxies is most successful for flare type prediction: the joint use of $\hcrit$ and $\Hjn$ (or $\Hjfn$) raises the success rate of flare type prediction in major flares from $\lesssim$70\% to over 90\%.
    
    The overall photospheric twist, $\alphat$, often used as an easy to compute proxy of the coronal helicity) is only moderately indicative for the amount of helicity present in the preflare corona (Pearson correlation $\lesssim$\,0.8). It is a slightly better proxy of the nonpotentiality than the total helicity. In addition, one likely guesses the sign of the helicity in the preflare corona wrong in at least 10\% (considering major flares only; see Fig.~\href{fig:jsoc_scatter}{\ref{fig:jsoc_scatter}}a). A flux-normalized measure of the flux-$R$ proxy (indicative of upcoming major flaring; originally introduced by \cite{2007ApJ...655L.117S}) we find as nonindicative of the type of upcoming major flaring and essentially uncorrelated with the preflare helicity-related proxies (Fig.~\href{fig:jsoc_scatter}{\ref{fig:jsoc_scatter}}c,d). 

\subsection{Flare-related changes}

    We provide observation-based model support for several aspects regarding the impact of major flares onto the coronal energy and helicity budgets of solar ARs (top panels in Fig.~\href{fig:histo_changes}{\ref{fig:histo_changes}}). First, the approximate conservation of $\Hv$ during magnetic reconnection. Second, this is also true even when there is a significant dissipation of magnetic energy. Third, flare-associated CMEs efficiently reduce the magnetic complexity of the corona ($\gtrsim$\,10\% when expressed in terms of the preflare total helicity). Fourth, energy is dissipated more efficiently during eruptive flares ($\gtrsim$\,10\% vs.\ $\sim$5\% during confined flares). 
    
    Although most strongly modified due to eruptivity, the complexity of the current-carrying field is also considerably altered in the course of confined flaring ($\approx$10\% with respect to the preflare level). This might partly be explained by helicity transfer to the volume-threading part \citep[][]{2018ApJ...865...52L}.
    
    In nearly 80\% of the major flares more than 20\% of the preflare content of $\Ef$ and $\Hv$ are simultaneously released, and the reductions being at least a factor of ten larger than characteristic time variations in the absence of major flaring. For smaller eruptive flares (GOES classes M1 to M4; bottom panels in Fig.\href{fig:histo_changes}{\ref{fig:histo_changes}}) we find eruptive-flare related reductions in $\approx$\,60\% of the events at most, and exceeding the magnitude of characteristic time variations (in the absence of major flaring) in only about 40\%. As a consequence, we suggest to question critically model-based assessments of flare-related changes for small flares.
    
    Characteristic flare-related changes to the total energy and helicity due to the occurrence of X-class flares amount to $\lesssim$15\% of the preflare content. In comparison, characteristic flare-related changes of the free energy and helicity of the current-carrying field are more pronounced ($\approx$\,20\% to 30\%). Despite the remaining budgets being large enough to actually fuel further major flares, the occurrence of eruptive X-class flares certainly represents a strong conditioning of the solar corona.

\subsection{Postflare evolution}

    The coronal energy and helicity budgets hardly return to near-preflare levels within $\sim$12~hours, on average, after eruptive X-class flares (left column in Fig.~\href{fig:mnsea_post}{\ref{fig:mnsea_post}}).
    We may now assume that the replenishment time serves as an indicator for the time required to (re)build a coronal configuration of sufficient magnetic complexity to produce an eruptive X-class flare (independently of the preflare configuration and, in our case, also in the absence of other major flaring). Then, we may have found a partial constraint to the ability of an AR to produce multiple eruptive X-class flares within a few hours. We identified only a single case in solar cycle~24 when this happened, namely the three X-class flares (of which two were eruptive) that originated from NOAA~10808 within a time window of $\sim$2.5~hours \citep[e.g.,][]{2009ApJ...703..757L}.
    
    Large eruptive M-class flares (GOES classes M5 to M9) affect the coronal budgets less strong with the replenishment of both energy and helicity budgets being accomplished within roughly 6 to 12~hours (right column of Fig.~\href{fig:mnsea_post}{\ref{fig:mnsea_post}}). 
    
    Smaller eruptive M-class flares (GOES classes M1 to M4) have only little impact on the coronal budgets, with the replenishment being accomplished in practically no time. This may be partly attributed to the limitations of NLFF modeling (for an in-depth discussion see Sect.~\href{ss:changes}{\ref{ss:changes}}).

\section{Summary and conclusions}
\label{s:summary}

We present an in-depth study of statistical aspects related to the time evolution of coronal quantities in association to large solar flaring (GOES class M1 or larger) and, in particular, to the type of flaring (confined or eruptive). We are interested in the coronal budgets of (free) magnetic energies and relative helicities of the total and nonpotential (current-carrying) field, as well as associated eruptivity proxies (relative measures), including the free-to-potential energy ratio and helicity ratio, as well as the flux-normalized helicities. To retrieve these quantities, we analyzed data-constrained NLFF model time series for 36 different solar ARs, where we chose optimization-based NLFF modeling as a tool and performed subsequent finite-volume relative helicity computation (cf., Sect.~\href{ss:modeling}{\ref{ss:modeling}}). For each of the employed NLFF models, the realism (as well as the numerical qualification for helicity computation) was explicitly tested and demonstrated  
(cf., Sect.~\href{ss:modeling}{\ref{ss:modeling}}). This is an important aspect for the validation of statistical studies such as this, but is often left out or is not included in a suitable way \citep[\eg,][]{2023ApJ...945..102D,2023RAA....23i5025W}.

The high-quality modeling allowed us to cover the coronal evolution around 231 solar flares  to investigate particular aspects, including (1) the long-term preflare time evolution, (2) the immediate preflare conditions, (3) flare-related changes, and (4) the postflare evolution. To address (1), (4), and partially (3), we synthesized information from selected major flare events using superposed epoch analysis \citep[][]{1913RSPTA.212...75C} with a higher statistical reliability compared to the pioneering study of \cite{2023ApJ...942...27L}. For the analysis of the relative time evolution of coronal quantities, we applied dynamic time warping \citep[\eg,][]{2001SIAM...1K} to the superposed epoch data. 

In summary, the study presented here provides detailed insights into the long-term evolution of the solar corona and its conditioning due to the occurrence of large flares. First, our study of the coronal conditions over the 48~hours prior to major flaring revealed that the flare type may possibly be indicated by the (more or less) pronounced similarity between the time evolution of specific photospheric and coronal quantities. For instance, during the 24~hours leading to a major confined flare, the free energy evolves more similar with respect to the unsigned photospheric flux than the helicity of the current-carrying field, while it is the opposite case prior to eruptive flares. We also found that CME-favorable conditions, when quantified via energy- and helicity-related eruptivity proxies, persist over extended time periods prior to eruptive major flaring. Therefore, the exceeding of related "critical values" may be used as an indicator for near-in-time upcoming eruptive flaring only with limited success. Second, based on our analysis of immediate preflare coronal conditions, we present evidence that the success in predicting the type of an upcoming major flare (\ie, whether or not it produces a CME) depends on the combination of different proxies, both regarding the local stability of a preeruptive structure as well as the nonpotentiality of the host AR. When using the critical height for torus instability and the normalized helicity of the current-carrying field as respective measures, the success rate of flare type prediction in major flares is more than 90\%. Third, from the analysis of flare-related changes we find the budgets of free energy and helicity of the current-carrying field to be significantly affected, which certainly represents a strong conditioning of the solar corona. Fourth, our study of the coronal conditions during the 12~hours after the occurrence of eruptive X-class flaring revealed that a time window of more than approximately 12~hours is required to replenish the coronal energy and helicity budgets. Assuming that such a time frame reflects the time generally required to build up a coronal configuration capable of fueling an eruptive X-class flare, our findings may serve as a partial explanation to the relative paucity of repeated eruptive X-class flaring from the same source AR.

\begin{acknowledgements}
We thank the anonymous referee for the valuable and constructive suggestions to improve our manuscript. This research was funded in part by the Austrian Science Fund (FWF) 10.55776/P31413. Y.\,L.\ acknowledges support from NASA LWS 80NSSC19K0072. J.\,K.\,T.\ acknowledges the High Performance Computing (HPC) center of the University of Graz for providing computational resources and technical support. NASA’s SDO satellite and the HMI instrument were joint efforts by many teams and individuals, whose efforts are greatly appreciated. We thank Dr.\ Robert Jarolim for providing the possibility to cross-check our flare-AR associations.
\end{acknowledgements}

\bibliographystyle{aa}
\bibliography{bibliography}

\begin{appendix}
\section{Flare sample}
\renewcommand{\thetable}{A\arabic{table}}
\setlength{\LTcapwidth}{\textwidth}
\scriptsize
\onecolumn
\begin{longtable}{>{\stepcounter{rowno}\therowno.}rllclllcc}
\caption{Large flares used in this work, including relevant information that supports the assigned flare type.
\label{tab:flare_summary}}\\
\hline\hline
\multicolumn{1}{r}{No.} & Flare identifier & Flare & NOAA & Flare & \multicolumn{2}{c}{CME-properties\tablefootmark{\href{f:lasco}{\ref{f:lasco}}}} & EUV  & Comments \\
\multicolumn{1}{r}{} & (start time) & position\tablefootmark{\href{f:ssle}{\ref{f:ssle}}} & number\tablefootmark{\href{f:ssle}{\ref{f:ssle}}} & type & $t_{\mathrm{first,C2}}$ (PA; $t_{0,\rm{linear}}$/$t_{0,\rm{quadratic}}$) & Speed (km\,s$^{-1}$) & signatures\tablefootmark{\href{f:sde}{\ref{f:sde}}} & \\
\hline
\endfirsthead
\caption{Continued.}\\
\hline\hline
\multicolumn{1}{r}{No.} & Flare identifier & Flare & NOAA & Flare & \multicolumn{2}{c}{CME-properties\tablefootmark{\href{f:lasco}{\ref{f:lasco}}}} & EUV  & Comments\\
\multicolumn{1}{r}{} & (start time) & position\tablefootmark{\href{f:ssle}{\ref{f:ssle}}} & number\tablefootmark{\href{f:ssle}{\ref{f:ssle}}} & type & $t_{\mathrm{first,C2}}$ (PA; $t_{0,\rm{linear}}$/$t_{0,\rm{quadratic}}$) & Speed (km\,s$^{-1}$) & signatures\tablefootmark{\href{f:sde}{\ref{f:sde}}} & \\
\hline
\endhead
\hline
\endfoot
\hline  
~ & \bf SOL2011-02-13T17:28M6.6 & S20E04 & 11158 & \cme\tablefootmark{b,c} & 18:36 (359; 17:53/16:10) & 373 & $+$ & ID 5 in Fig.~\ref{fig:dtw_major}\\
~ & SOL2011-02-14T17:26M2.2 & S20W04\tablefootmark{\href{f:hfc}{\ref{f:hfc}}} & 11158 & \cme$^{\rm(h)}$ & 18:24 (HALO; 17:18/17:08) & 326 & $+$ & \\
~ & \bf SOL2011-02-15T01:44X2.2 & S20W10 & 11158 & \cme\tablefootmark{a,b,c} & 02:24 (HALO; 01:49/02:04) & 669 & $+$ & ID 6 in Fig.~\ref{fig:dtw_major}\\
~ & SOL2011-02-16T01:32M1.0 & S20W24\tablefootmark{\href{f:hfc}{\ref{f:hfc}}} & 11158 & C & $\times$ & $\times$ & $-$ & \\ 
~ & SOL2011-02-16T07:35M1.1 & S20W30 & 11158 & C\tablefootmark{c} & $\times$ & $\times$ & $-$ & \\
~ & SOL2011-02-16T14:19M1.6 & S20W32 & 11158 & E, \gray C\tablefootmark{c} & $\times$ & $\times$ & $+$ & \\
\hline
~ & SOL2011-03-07T13:45M1.7 & N11E27 & 11166 & \cme\tablefootmark{c} & 14:48 (354; 14:06/14:15) & 689 & $+$ & \\
~ & SOL2011-03-09T10:35M1.7 & N11W01 & 11166 & \cme\tablefootmark{c} & 12:12 (238; 11:04/10:19) & 315 & $+$ & \\
~ & SOL2011-03-09T13:17M1.7 & N09W06 & 11166 & C\tablefootmark{c} & $\times$ & $\times$ & $-$ & \\
~ & \bf SOL2011-03-09T23:13X1.5 & N08W09 & 11166 & C\tablefootmark{a,b,c} & $\times$ & $\times$ & $-$ & ID 7 in Fig.~\ref{fig:dtw_major}\\
~ & SOL2011-03-10T22:34M1.1 & N08W25\tablefootmark{\href{f:hfc}{\ref{f:hfc}}} & 11166 & C & $\times$ & $\times$ & $-$ & \\ 
\hline
~ & \bf SOL2011-07-30T02:04M9.3 & N14E35 & 11261 & C\tablefootmark{b} & $\times$ & $\times$ & $-$ & ID 8 in Fig.~\ref{fig:dtw_major}\\
~ & SOL2011-08-02T05:19M1.4 & N16W08 & 11261 & \cme\tablefootmark{c} & 06:36 (288; 05:57/06:12) & 712 & $+$ & \\
~ & SOL2011-08-03T03:08M1.1 & N17W24 & 11261 & C\tablefootmark{c} & $\times$ & $\times$ & $-$ & \\
~ & \bf SOL2011-08-03T13:17M6.0 & N16W30 & 11261 & \cme\tablefootmark{b,c} & 14:00 (HALO; 13:06/13:22) & 610 & $+$ & ID 9 in Fig.~\ref{fig:dtw_major}\\
~ & \bf SOL2011-08-04T03:41M9.3 & N19W36 & 11261 & \cme\tablefootmark{b,c} & 04:12 (HALO; 03:39/03:47) & 1315 & $+$ & \\
\hline
~ & \bf SOL2011-09-06T01:35M5.3 & N14W07 & 11283 & \cme\tablefootmark{b,c} & 02:14 (HALO; 02:00/01:04) & 782 & $+$ & ID 10 in Fig.~\ref{fig:dtw_major}\\
~ & \bf SOL2011-09-06T22:12X2.1 & N14W18 & 11283 & \cme\tablefootmark{a,b,c} & 23:05 (HALO; 21:58/21:55) & 575 & $+$ & \\
~ & \bf SOL2011-09-07T22:32X1.8 & N14W28 & 11283 & \cme\tablefootmark{a,b,c} & 23:05 (290; 22:34/22:35) & 792 & $+$ & ID 11 in Fig.~\ref{fig:dtw_major}\\
~ & \bf SOL2011-09-08T15:32M6.7 & N14W40 & 11283 & \cme\tablefootmark{b}, \gray C\tablefootmark{c} & 16:36 (317; 15:11/13:47) & 214 & $+$ & \\
\hline
~ & SOL2011-09-25T15:26M3.7 & N16E43 & 11302 & \cme\tablefootmark{c} & 16:00 (107; 15:07/15:23) & 676 & $+$ & \\
~ & SOL2011-09-25T16:51M2.2 & N12E41\tablefootmark{\href{f:hfc}{\ref{f:hfc}}} & 11302 & C & $\times$ & $\times$ & $-$ & \\
~ & SOL2011-09-26T05:06M4.0 & N13E34 & 11302 & C & $\times$ & $\times$ & $-$ & \\
~ & SOL2011-09-26T14:37M2.6 & N14E30 & 11302 & \cme\tablefootmark{c} & 15:12 (86; 14:03/13:19) & 240 & $+$ & \\
\hline
~ & SOL2011-09-30T18:55M1.0 & N08E06 & 11305 & \cme\tablefootmark{c} & 20:00 (77; 19:13/18:05) & 337 & $+$ & \\
~ & SOL2011-10-01T08:56M1.2 & N10W06 & 11305 & \cme\tablefootmark{c} & 09:36 (317; 08:36/08:45) & 448 & $+$ &  \\
~ & SOL2011-10-02T00:37M3.9 & N09W12 & 11305 & \cme\tablefootmark{c} & 02:00 (167; 00:42/00:40) & 259 & $+$ & \\
~ & SOL2011-11-05T20:31M1.8 & N21E34 & 11339 & C\tablefootmark{c} & $\times$ & $\times$ & $-$ & \\ 
~ & SOL2011-11-06T00:46M1.2 & N21E35 & 11339 & C\tablefootmark{c} & $\times$ & $\times$ & $-$ & \\ 
~ & SOL2011-11-06T06:14M1.4 & N21E31 & 11339 & C\tablefootmark{c} & $\times$ & $\times$ & $-$ & \\ 
\hline
\hline
~ & SOL2012-03-06T00:22M1.3 & N16E41 & 11429 & C\tablefootmark{c} & $\times$ & $\times$ & $-$ & \\
~ & SOL2012-03-06T01:36M1.2 & N17E41 & 11429 & C\tablefootmark{c} & $\times$ & $\times$ & $-$ & \\
~ & SOL2012-03-06T04:01M1.0 & N16E39 & 11429 & E, \gray C\tablefootmark{c} & 04:48 (32; 05:01/03:49) & 536 & $+$ & \\
~ & SOL2012-03-06T07:52M1.0 & N17E41 & 11429 & \cme\tablefootmark{c} & 08:12 (34; 07:45/07:38) & 599 & $+$ & \\
~ & SOL2012-03-06T12:23M2.1 & N18E36 & 11429 & C\tablefootmark{c} & $\times$ & $\times$ & $-$ & \\
~ & SOL2012-03-06T21:04M1.3 & N16E30\tablefootmark{\href{f:hfc}{\ref{f:hfc}}} & 11429 & C & $\times$ & $\times$ & $-$ & \\
~ & SOL2012-03-06T22:49M1.0 & N17E35 & 11429 & C\tablefootmark{c} & $\times$ & $\times$ & $-$ & \\
~ & \bf SOL2012-03-07T00:02X5.4 & N17E31 & 11429 & \cme\tablefootmark{a,b,c} & 00:24 (HALO; 00:16/00:18) & 2684 & $+$ & ID 12 in Fig.~\ref{fig:dtw_major}\\
~ & \bf SOL2012-03-07T01:05X1.3 & N17E20 & 11429 & \cme\tablefootmark{a,b,c} & 01:30 (HALO; 00:56/01:04) & 1825 & $+$ & \\
~ & \bf SOL2012-03-09T03:22M6.3 & N15W03 & 11429 & \cme\tablefootmark{b,c} & 04:26 (HALO; 03:43/03:50) & 950 & $+$ & ID 13 in Fig.~\ref{fig:dtw_major}\\
~ & \bf SOL2012-03-10T17:15M8.4 & N17W24 & 11429 & \cme\tablefootmark{b} & 18:00 (HALO; 17:35/17:37) & 1296 & $+$ & ID 14 in Fig.~\ref{fig:dtw_major}\\
\hline
~ & SOL2012-05-09T12:21M4.7 & N13E31 & 11476 & C\tablefootmark{c} & $\times$ & $\times$ & $-$ & \\
~ & SOL2012-05-09T14:02M1.8 & N06E22 & 11476 & C\tablefootmark{c} & $\times$ & $\times$ & $-$ & \\
~ & SOL2012-05-09T21:01M4.1 & N11E25 & 11476 & \cme\tablefootmark{c} & $\times$ & $\times$ & $+$ & \\ 
~ & \bf SOL2012-05-10T04:11M5.7 & N13E22 & 11476 & C\tablefootmark{c} & $\times$ & $\times$ & $-$ & ID 15 in Fig.~\ref{fig:dtw_major}\\
~ & SOL2012-05-10T20:20M1.7 & N10E12 & 11476 & C\tablefootmark{c} & $\times$ & $\times$ & $-$ & \\
\hline
 ~ & SOL2012-06-30T12:48M1.0 & N17E21 & 11513 & C & $\times$ & $\times$ & $-$ & \\
 ~ & SOL2012-06-30T18:26M1.6 & N17E20 & 11513 & E & 18:48 (68; 17:38/17:03) & 247 & $+$ & \\
 ~ & SOL2012-07-01T19:11M2.8 & N14E04 & 11513 & C & $\times$ & $\times$ & $-$ &  \\
 ~ & SOL2012-07-02T00:26M1.1 & N15E01 & 11513 & C & $\times$ & $\times$ & $-$ &  \\
 ~ & SOL2012-07-04T16:33M1.8 & N14W34 & 11513 & E & 17:24 (HALO; 16:50/16:57) & 662 & $+$ & \\
\hline
~ & \bf SOL2012-07-02T10:34M5.6 & S17E01 & 11515 & \cme\tablefootmark{b,c} & 11:24 (174; 09:56/10:22) & 313 & $+$ & ID 16 in Fig.~\ref{fig:dtw_major}\\
~ & SOL2012-07-02T19:59M3.8 & S17W01 & 11515 & \cme\tablefootmark{c} & 20:24 (185; 19:14/19:51) & 527 & $+$ & \\
~ & SOL2012-07-02T23:49M2.0 & S16W02 & 11515 & \cme\tablefootmark{c} & +00:48 (195; +00:34/23:12) & 400 & $+$ & \\
~ & SOL2012-07-04T04:28M2.3 & S17W18 & 11515 & \cme\tablefootmark{c} & 05:12 (209; 04:27/04:00) & 381 & $-$ & \\
~ & \bf SOL2012-07-04T09:47M5.3 & S20W18 & 11515 & C, \gray\cme\tablefootmark{c} & $\times$ & $\times$ & $-$ & ID 17 in Fig.~\ref{fig:dtw_major}\\
~ & SOL2012-07-04T12:07M2.3 & S16W18 & 11515 & \cme\tablefootmark{c} & 12:48 (203; 11:14/11:50) & 290 & $+$ & \\
~ & SOL2012-07-04T14:35M1.3 & S18W18 & 11515 & C; \gray \cme\tablefootmark{c} & $\times$ & $\times$ & $-$ & \\
~ & SOL2012-07-04T22:03M4.6 & S16W23 & 11515 & \cme\tablefootmark{c} & 22:36 (195; 21:38/21:57) & 556 & $+$ & \\
~ & SOL2012-07-04T23:47M1.3 & S16W21 & 11515 & C\tablefootmark{c} & $\times$ & $\times$ & $-$ & \\
~ & SOL2012-07-05T01:05M2.4 & S18W26 & 11515 & C\tablefootmark{c} & $\times$ & $\times$ & $-$ & \\
~ & SOL2012-07-05T02:35M2.2 & S17W23 & 11515 & C\tablefootmark{c} & $\times$ & $\times$ & $\times$ &  \\ 
~ & SOL2012-07-05T03:25M4.7 & S17W23 & 11515 & C\tablefootmark{c} & $\times$ & $\times$ & $-$ & \\ 
~ & SOL2012-07-05T06:49M1.1 & S18W39 & 11515 & \cme\tablefootmark{c} & 08:00 (320; 06:59/06:34) & 329 & $+$ & \\
~ & SOL2012-07-05T10:44M1.8 & S19W30 & 11515 & C\tablefootmark{c} & $\times$ & $\times$ & $-$ & \\
~ & \bf SOL2012-07-05T11:39M6.1 & S22W32 & 11515 & C\tablefootmark{c} & $\times$ & $\times$ & $-$ & ID 18 in Fig.~\ref{fig:dtw_major}\\
~ & SOL2012-07-05T13:05M1.2 & S16W43 & 11515 & \cme\tablefootmark{c} & 13:24 (219; 12:50/12:48) & 741 & $+$ & \\
~ & SOL2012-07-05T20:09M1.6 & S17W35 & 11515 & C; \gray \cme\tablefootmark{c} & $\times$ & $\times$ & $-$ & \\
~ & SOL2012-07-06T01:37M2.9 & S18W41 & 11515 & C\tablefootmark{c} & $\times$ & $\times$ & $-$ & \\
~ & SOL2012-07-06T02:44M1.0 & S17W40 & 11515 & \cme\tablefootmark{c} & 03:12 (228; 02:24/02:44) & 1059 & $\times$ &  \\
\hline
~ & SOL2012-07-09T23:03M1.1 & S15E43 & 11520 & C\tablefootmark{c} & $\times$ & $\times$ & $-$ & \\
~ & SOL2012-07-10T04:58M1.7 & S17E33 & 11520 & C\tablefootmark{c} & $\times$ & $\times$ & $-$ & \\ 
~ & SOL2012-07-10T06:05M2.0 & S17E30 & 11520 & C\tablefootmark{c} & $\times$ & $\times$ & $-$ & \\
~ & \bf SOL2012-07-12T15:37X1.4 & S15W01 & 11520 & \cme\tablefootmark{a,b,c} & 16:24 (228; 15:56/16:02) & 843 & $-$ & ID 19 in Fig.~\ref{fig:dtw_major}\\
\hline
~ & SOL2012-11-20T19:21M1.6 & N07E15 & 11618 & \cme\tablefootmark{c} & $\times$ & $\times $& $+$ & \\ 
~ & SOL2012-11-21T06:45M1.4 & N06E10 & 11618 & \cme\tablefootmark{c} & 08:37 (63; {\gray 08:03}/07:35) & 410 & $+$ & \\
~ & SOL2012-11-21T15:10M3.5 & N08E14 & 11618 & \cme\tablefootmark{c} & 16:00 (HALO; 14:55/15:13) & 529 & & \\
\hline
~ & \bf SOL2013-04-11T06:55M6.5 & N09E12 & 11719 & \cme\tablefootmark{c} & 07:24 (HALO; 06:50/06:55) & 861 & $+$ & ID 20 in Fig.~\ref{fig:dtw_major}\\
\hline
~ & SOL2013-05-17T08:43M3.2 & N10E35 & 11748 & \cme\tablefootmark{c} & 09:12 (HALO; 08:46/08:47) & 1345 & $+$ & \\
\hline

\hline
~ & SOL2013-10-13T00:12M1.7 & S22E17 & 11865 & \cme\tablefootmark{c} & 01:25 (146; 00:33/00:42) & 478 & $+$ & \\
~ & SOL2013-10-15T08:26M1.8 & S22W13 & 11865 & \cme\tablefootmark{c} & 09:36 (196; 08:08/08:38) & 223 & $+$ & \\
\hline
~ & SOL2013-10-22T00:14M1.0 & N06E17 & 11875 & C\tablefootmark{c} & $\times$ & $\times$ & $-$ & \\ 
~ & SOL2013-10-22T14:49M1.0 & N07E07 & 11875 & E, \gray C\tablefootmark{c} & 15:24 (82; 14:32/14:46) & 351 & $-$ & \\ 
~ & SOL2013-10-22T21:15M4.2 & N04W01 & 11875 & \cme\tablefootmark{c} & 21:48 (278; 21:09/21:14) & 589 & $+$ &  \\
~ & SOL2013-10-23T20:41M2.7 & N07W04 & 11875 & C\tablefootmark{c} & $\times$ & $\times$ & $-$ & \\ 
~ & SOL2013-10-23T23:33M1.4 & N07W07 & 11875 & C\tablefootmark{c} & $\times$ & $\times$ & $-$ & \\ 
~ & SOL2013-10-23T23:58M3.1 & N08W11 & 11875 & C\tablefootmark{c} &  $\times$ & $\times$ & $-$ & \\ 
~ & SOL2013-10-24T09:59M2.5 & N06W14 & 11875 & C\tablefootmark{c} & $\times$ & $\times$ & $-$ & \\ 
~ & SOL2013-10-24T10:30M3.5 & N06W12 & 11875 & C & $\times$ & $\times$ & $-$ & \\ 
\hline
~ & \bf SOL2013-11-05T22:07X3.3 & S12E44 & 11890 & \cme\tablefootmark{c} & 22:26 (160; 21:36/21:48) & 562 & $+$ & ID 21 in Fig.~\ref{fig:dtw_major}\\
~ & SOL2013-11-06T13:39M3.8 & S11E36 & 11890 & \cme\tablefootmark{c} & 14:24 (147; 13:23/13:40) & 347 & $+$ & \\
~ & SOL2013-11-07T03:34M2.3 & S14E28 & 11890 & \cme\tablefootmark{c} & 04:24 (177; 03:37/03:46) & 373 & $+$ & \\
~ & SOL2013-11-07T14:15M2.4 & S13E23 & 11890 & \cme\tablefootmark{c} & 15:12 (HALO; 14:00/14:12) & 411 & $+$ & \\
~ & \bf SOL2013-11-08T04:20X1.1 & S12E13 & 11890 & \cme\tablefootmark{a,b,c} & 03:24 (HALO; 03:23/02:34) & 497 & $+$ & ID 22 in Fig.~\ref{fig:dtw_major}\\ 
~ & \bf SOL2013-11-10T05:08X1.1 & S14W13 & 11890 & \cme\tablefootmark{a,b,c} & 05:36 (220; 04:55/05:15) & 682 & $+$ & ID 23 in Fig.~\ref{fig:dtw_major} \\
\hline
~ & SOL2013-12-29T07:49M3.1 & S16W01 & 11936 & C, \gray E\tablefootmark{c} & $\times$ & $\times$ & $-$ & \\        
~ & \bf SOL2013-12-31T21:45M6.4 & S15W36 & 11936 & E\tablefootmark{c} & 22:36 (237; 12:25/12:01) & 271 & $+$ & ID 24 in Fig.~\ref{fig:dtw_major} \\   
\hline
~ & SOL2014-01-31T15:32M1.1 & N09E36 & 11968 & \cme\tablefootmark{c} & 16:24 (8; 15:34/15:52) & 462 & $+$ & \\
~ & SOL2014-02-02T06:24M2.6 & N12E18 & 11968 & \cme\tablefootmark{c} & 06:48 (94; 05:32/05:29) & 230 & $+$ & \\
~ & SOL2014-02-02T14:01M1.3 & N12E18 & 11968 & \cme\tablefootmark{c} & $\times$ & $\times$ & $+$ & \\
~ & SOL2014-02-02T16:24M1.0 & N10E05 & 11968 & C\tablefootmark{c} & $\times$ & $\times$ & $-$ & \\ 
\hline
~ & SOL2014-02-01T01:19M1.0 & S11E26 & 11967 & C\tablefootmark{c} & $\times$ & $\times$ & $-$ & \\ 
~ & SOL2014-02-01T07:14M3.0 & S11E23 & 11967 & C\tablefootmark{c} & $\times$ & $\times$ & $-$ & \\ 
~ & SOL2014-02-02T07:17M2.2 & S10E14 & 11967 & C\tablefootmark{c} & $\times$ & $\times$ & $-$ & \\ 
~ & SOL2014-02-02T09:24M4.4 & S11E13 & 11967 & C\tablefootmark{c} & $\times$ & $\times$ & $-$ & \\ 
~ & SOL2014-02-02T18:05M3.1 & S13E05 & 11967 & C\tablefootmark{c} & $\times$ & $\times$ & $-$ & \\ 
~ & SOL2014-02-02T21:24M1.3 & S13E05 & 11967 & C\tablefootmark{c} & \gray 23:48 (120; 22:22/22:35) & \gray 199 & $-$ & \\ 
~ & SOL2014-02-04T01:16M3.8 & S13W14 & 11967 & C\tablefootmark{c} & $\times$ & $\times$ & $-$ & \\ 
~ & \bf SOL2014-02-04T03:57M5.2 & S14W06 & 11967 & C\tablefootmark{c} & $\times$ & $\times$ & $-$ & ID 25 in Fig.~\ref{fig:dtw_major} \\
~ & SOL2014-02-04T09:38M1.4 & S13W12 & 11967 & C\tablefootmark{c} & $\times$ & $\times$ & $-$ & \\ 
~ & SOL2014-02-04T15:25M1.5 & S12W12 & 11967 & \cme\tablefootmark{c} & 16:36 (250; 15:30/15:40) & 368 & $+$ & \\
\hline
~ & SOL2014-02-11T03:22M1.7 & S12E17 &  11974 & \cme\tablefootmark{c} & \gray $\times$ & \gray $\times$ & $+$ & \\ 
~ & SOL2014-02-11T16:34M1.8 & S13E12 &  11974 & \cme\tablefootmark{c} & $\times$ & $\times$ & $+$ & \\
~ & SOL2014-02-12T03:52M3.7 & S12W02 &  11974 & \cme\tablefootmark{c} & $\times$ & $\times$ & $+$ & \\
~ & SOL2014-02-12T06:54M2.3 & S12E01 &  11974 & \cme\tablefootmark{c} & 08:12 (136; 08:59/06:59) & 274 & $-$ & \\
~ & SOL2014-02-13T01:32M1.8 & S12W12 &  11974 & C\tablefootmark{c} & $\times$ & $\times$ &  $-$ & \\
~ & SOL2014-02-13T02:41M1.0 & S12W12 &  11974 & \cme\tablefootmark{c} & $\times$ & $\times$ &  $-$ & \\
~ & SOL2014-02-13T05:49M1.7 & S12W12 &  11974 & C\tablefootmark{c} & $\times$ & $\times$ &  $-$ & \\
~ & SOL2014-02-13T08:05M1.0 & S12W13 &  11974 & C\tablefootmark{c} & $\times$ & $\times$ &  $-$ & \\
~ & SOL2014-02-13T15:45M1.4 & S12W29 &  11974 & \cme\tablefootmark{c} & 16:36 (205; 15:31/15:57) & 502 &  $-$ & \\
~ & SOL2014-02-14T02:40M2.3 & S12W25 &  11974 & \cme\tablefootmark{c} & \gray $\times$ & \gray $\times$ & $+$ & \\ 
~ & SOL2014-02-14T12:29M1.6 & S15W36 &  11974 & \cme\tablefootmark{c} & $\times$ & $\times$ & $+$ & \\
~ & SOL2014-02-14T13:21M1.1 & S12W30 &  11974 & C\tablefootmark{c} & $\times$ & $\times$ & $-$ & \\
~ & SOL2014-02-14T16:33M1.0 & S12W32 &  11974 & \cme\tablefootmark{c} & 17:24 (252; 16:24/16:21) & 283 & $+$ & \\
\hline
~ & SOL2014-03-28T19:04M2.0 & N11W21 & 12017 & \cme\tablefootmark{c} & $\times$ & $\times$ & \\ 
~ & SOL2014-03-28T23:44M2.6 & N10W22 & 12017 & \cme\tablefootmark{c} & 23:48 (325; 22:55/23:06) & 514 & $+$ & \\
~ & \bf SOL2014-03-29T17:35X1.0 & N10W32 & 12017 & \cme\tablefootmark{a,b,c} & 18:12 (HALO; 17:12/17:27) & 528 & $+$ & ID 1 in Fig.~\ref{fig:dtw_major}\\
~ & SOL2014-03-30T11:48M2.1 & N08W43 & 12017 & \cme\tablefootmark{c} & 12:14 (291; 11:28/11:37) & 487 & $+$& \\
\hline
~ & SOL2014-04-18T12:31M7.3 & S20W34 & 12036 & E\tablefootmark{c} & 13:25 (HALO; 12:43/12:38) & 1203 & $+$ & ID 2 in Fig.~\ref{fig:dtw_major}\\
\hline
~ & SOL2014-06-13T07:49M2.6 & S18E40 & 12087 & \cme\tablefootmark{c} & 08:24 (127; 07:17/07:33) & 370 & $+$ & \\
~ & SOL2014-06-15T23:50M1.0 & S22E07 & 12087 & \cme\tablefootmark{c} & +01:00 (\textcolor{gray}{228}; +00:03/23:50) & 347 & $+$ & \\ 
\hline
~ & SOL2014-09-08T23:12M4.6 & N14E31 & 12158 & E & 00:06 (HALO; 23:45/23:49) & 920 & $+$ & \\
~ & SOL2014-09-10T17:21X1.6 & N11E05 & 12158 & E &  18:00 (HALO; 17:27/17:37) & 1267 & $+$ & ID 26 in Fig.~\ref{fig:dtw_major} \\
\hline
~ & SOL2014-10-20T16:00M4.5 & S14E37 & 12192 & C\tablefootmark{c} & $\times$ & $\times$ & $-$ & \\ 
~ & SOL2014-10-20T18:55M1.4 & S13E43 & 12192 & C\tablefootmark{c} & $\times$ & $\times$ & $-$ & \\
~ & SOL2014-10-20T19:53M1.7 & S13E43 & 12192 & C\tablefootmark{c} & $\times$ & $\times$ & $-$ & \\
~ & SOL2014-10-20T22:43M1.2 & S14E36 & 12192 & C\tablefootmark{c} & $\times$ & $\times$ & $-$ & \\ 
~ & SOL2014-10-21T13:35M1.2 & S14E35\tablefootmark{\href{f:hfc}{\ref{f:hfc}}} & 12192 & C & $\times$ & $\times$ & $-$ & \\
~ & \bf SOL2014-10-22T01:06M8.7 & S12E21 & 12192 & C\tablefootmark{b,c} & $\times$ & $\times$ & $-$ & ID 27 in Fig.~\ref{fig:dtw_major} \\ 
~ & SOL2014-10-22T05:11M2.7 & S14E19 & 12192 & C\tablefootmark{c} & $\times$ & $\times$ & $-$ & \\ 
~ & \bf SOL2014-10-22T14:02X1.6 & S14E13 & 12192 & C\tablefootmark{a,b,c} & $\times$ & $\times$ & $-$ &\\
~ & SOL2014-10-23T09:44M1.1 & S16E03 & 12192 & C\tablefootmark{c} & $\times$ & $\times$ & $-$ & \\ 
~ & SOL2014-10-24T07:37M4.0 & S19W06 & 12192 & \cme\tablefootmark{c} & 08:00 (215; 07:28/07:40) & 677 & $+$ & \\
~ & \bf SOL2014-10-24T21:07X3.1 & S22W21 & 12192 & C\tablefootmark{a,b,c} & $\times$ & $\times$ & $-$ & ID 28 in Fig.~\ref{fig:dtw_major} \\ 
~ & \bf SOL2014-10-25T16:55X1.0 & S10W22 & 12192 & C\tablefootmark{a,b,c} & $\times$ & $\times$ & $-$ & \\ 
~ & \bf SOL2014-10-26T10:04X2.0 & S14W37 & 12192 & C\tablefootmark{a,b,c} & $\times$ & $\times$ & $-$ & \\ 
~ & SOL2014-10-26T17:08M1.0 & S16W36 & 12192 & C\tablefootmark{c} & $\times$ & $\times$ & $-$ & \\  
~ & SOL2014-10-26T18:07M4.2 & S16W34 & 12192 & C\tablefootmark{c} & $\times$ & $\times$ & $-$ & \\  
~ & SOL2014-10-26T18:43M1.9 & S16W38 & 12192 & C\tablefootmark{c} & $\times$ & $\times$ & $-$ & \\  
~ & SOL2014-10-26T19:59M2.4 & S16W40 & 12192 & C\tablefootmark{c} & $\times$ & $\times$ & $-$ & \\ 
~ & SOL2014-10-27T00:06M7.1 & S14W44 & 12192 & C\tablefootmark{c} & $\times$ & $\times$ & $-$ & \\ 
~ & SOL2014-10-27T01:44M1.0 & S13W45 & 12192 & C\tablefootmark{c} & $\times$ & $\times$ & $-$ & \\ 
\hline
~ & \bf SOL2014-11-07T16:53X1.6 & N17E40 & 12205 & \cme\tablefootmark{a,b,c} & 17:12 (79; 16:28/16:49) & 469 & $+$ &\\
 ~ & SOL2014-11-09T15:24M2.3 & N18E14 & 12205 & \cme\tablefootmark{c} & 16:24 (\textcolor{gray}{307}; 15:36/14:57) & 388 & $+$ &  \\ 
\hline
~ & SOL2014-12-01T06:26M1.8 & S21E17 & 12222 & C\tablefootmark{c} & $\times$ & $\times$ & $-$ & \\ 
~ & SOL2014-12-04T08:00M1.3 & S24W27 & 12222 & C\tablefootmark{c} & $\times$ & $\times$ & $-$ & \\ 
~ & \bf SOL2014-12-04T18:05M6.1 & S20W31 & 12222 & C\tablefootmark{b} & $\times$ & $\times$ & $-$ & ID 29 in Fig.~\ref{fig:dtw_major} \\
~ & SOL2014-12-04T19:38M1.3 & S20W32 & 12222 & C\tablefootmark{c} & $\times$ & $\times$ & $-$ & \\ 
~ & SOL2014-12-05T11:33M1.5 & S19W37 & 12222 & C\tablefootmark{c} & $\times$ & $\times$ & $-$ & \\ 
\hline
~ & SOL2014-12-14T19:25M1.6 & S19E44 & 12242 & \cme\tablefootmark{c} & 19:48 (126; 19:05/19:10) & 626 & $+$ & \\
~ & SOL2014-12-17T00:57M1.5 & S25E10 & 12242 & C\tablefootmark{c} & $\times$ & $\times$ & \gray $+$ & \\ 
~ & \bf SOL2014-12-17T04:25M8.7 & S18E08 & 12242 & \cme\tablefootmark{b,c} & 05:00 (HALO; 04:01/04:06) & 587 & $+$ & ID 3 in Fig.~\ref{fig:dtw_major} \\
~ & SOL2014-12-19T09:31M1.3 & S19W27 & 12242 & C\tablefootmark{c} & $\times$ & $\times$ & \gray $+$ & \\ 
~ & \bf SOL2014-12-20T00:11X1.8 & S19W29 & 12242 & \cme\tablefootmark{b,c} & 01:25 (216; 00:13/00:25) & 830 & $+$ & ID 4 in Fig.~\ref{fig:dtw_major} \\
\hline
~ & SOL2014-12-17T01:41M1.1 & S11E33 & 12241 & \cme\tablefootmark{c} & 02:00 (107; 01:36/01:22) & 869 & $+$ & \\
~ & SOL2014-12-17T18:54M1.4 & S10E24 & 12241 & C\tablefootmark{c} & $\times$ & $\times$ & $-$ & \\ 
~ & \bf SOL2014-12-18T21:41M6.9 & S15E08 & 12241 & \cme\tablefootmark{c} & +01:04 (HALO; 22:02/22:48) & 1195 & $+$ & ID 30 in Fig.~\ref{fig:dtw_major} \\
~ & SOL2014-12-21T11:24M1.0 & S11W21 & 12241 & \cme\tablefootmark{c} & 12:12 (HALO; 11:34/11:41) & 669 & $+$ & \\
\hline
~ & SOL2015-01-26T16:46M1.1 & S10E25 & 12268 & C\tablefootmark{c} & $\times$ & $\times$ & $-$ & \\
~ & SOL2015-01-28T04:21M1.4 & S09E09 & 12268 & C\tablefootmark{c} & \gray 06:36 (110; 05:35/04:49) & \gray 321 & $-$ & \\ 
~ & SOL2015-01-29T11:32M2.1 & S12W06 & 12268 & C\tablefootmark{c} & $\times$ & $\times$ & $-$ & \\ 
~ & SOL2015-01-30T00:32M2.0 & S10W17 & 12268 & C\tablefootmark{c} & $\times$ & $\times$ & $-$ & \\ 
~ & SOL2015-01-30T05:29M1.7 & S10W17 & 12268 & C\tablefootmark{c} & $\times$ & $\times$ & $-$ & \\ 
\hline
~ & \bf SOL2015-03-10T03:19M5.1 & S15E39 & 12297 & \cme\tablefootmark{b} & 03:36 (HALO; 03:01/03:11) & 1040 & $+$ & \\
~ & SOL2015-03-10T23:46M2.9 & S16E28 & 12297 & \cme\tablefootmark{c} & +00:24 (81; 23:48/23:51) & 702 & $+$ & \\ 
~ & SOL2015-03-11T07:10M1.8 & S16E26\tablefootmark{\href{f:hfc}{\ref{f:hfc}}} & 12297 & C & $\times$ & $\times$ & $-$ & \\ 
~ & SOL2015-03-11T07:51M2.6 & S15E23\tablefootmark{\href{f:hfc}{\ref{f:hfc}}} & 12297 & C & $\times$ & $\times$ & $-$ & \\ 
~ & \bf SOL2015-03-11T16:11X2.1 & S12E22 & 12297 & \cme\tablefootmark{c}, \gray C\tablefootmark{b} & 17:00 (73; 14:48/16:02) & 240 & $+$ & ID 31 in Fig.~\ref{fig:dtw_major} \\
~ & SOL2015-03-11T18:37M1.0 & S16E18 & 12297 & C; \gray \cme\tablefootmark{c} & $\times$ & $\times$ & $-$ & \\
~ & SOL2015-03-12T04:41M3.2 & S16E14 & 12297 & C, \gray \cme\tablefootmark{c} & $\times$ & $\times$ & $-$ & \\
~ & SOL2015-03-12T11:38M1.6 & S16E14 & 12297 & C\tablefootmark{c} & $\times$ & $\times$ & $-$ & \\
~ & SOL2015-03-12T12:09M1.4 & S16E06 & 12297 & E, \gray C\tablefootmark{c} & \gray $\times$ & \gray $\times$ & $+$ & \\ 
~ & SOL2015-03-12T13:50M4.2 & S15E06 & 12297 & C\tablefootmark{c} & $\times$ & $\times$ & $-$ & \\
~ & SOL2015-03-12T21:44M2.7 & S16E04 & 12297 & E, \gray C\tablefootmark{c} & \gray $\times$ & \gray $\times$ & $+$ & \\ 
~ & SOL2015-03-13T03:47M1.2 & S17E03\tablefootmark{\href{f:hfc}{\ref{f:hfc}}} & 12297 & C & $\times$ & $\times$ & $-$ & \\ 
~ & SOL2015-03-13T05:49M1.8 & S14W02\tablefootmark{\href{f:hfc}{\ref{f:hfc}}} & 12297 & C & $\times$ & $\times$ & $-$ & \\
~ & SOL2015-03-14T04:23M1.3 & S17W13 & 12297 & C\tablefootmark{c} & $\times$ & $\times$ & $-$ & \\
~ & SOL2015-03-15T09:36M1.0 & S17W25 & 12297 & C\tablefootmark{c} & $\times$ & $\times$ & $-$ & \\
\hline
~ & SOL2015-06-20T06:28M1.0 & N13E27 & 12371 & \cme\tablefootmark{b,c} & 07:36 (120; 06:51/07:04) & 435 & $+$ & \\
~ & SOL2015-06-21T01:02M2.0 & N12E13 & 12371 & \cme\tablefootmark{b,c} & $\times$ & $\times$ & $+$ & \\
~ & SOL2015-06-21T02:06M2.6 & N13E14 & 12371 & \cme\tablefootmark{b,c} & 02:36 (HALO; 02:15/02:10) & 1366 & $+$ & \\
~ & \bf SOL2015-06-22T17:39M6.5 & N13W06 & 12371 & \cme\tablefootmark{b,c} & 18:36 (HALO; 17:58/18:06) & 1209 & $+$ & ID 32 in Fig.~\ref{fig:dtw_major} \\
~ & \bf SOL2015-06-25T08:02M7.9 & N12W40 & 12371 & \cme\tablefootmark{b,c} & 08:36 (HALO; 08:17/08:21) & 1627 & $+$ & ID 33 in Fig.~\ref{fig:dtw_major} \\
\hline
~ & SOL2015-08-21T01:56M1.2 & S16E39 & 12403 & C\tablefootmark{c} & $\times$ & $\times$ & $-$ & \\
~ & SOL2015-08-21T09:34M1.4 & S17E26 & 12403 & \cme\tablefootmark{c} & 10:12 (131; 09:19/09:05) & 555 & $+$ & \\
~ & SOL2015-08-21T19:10M1.1 & S12E26 & 12403 & \cme\tablefootmark{c} & $\times$ & $\times$ & $-$ & \\
~ & SOL2015-08-22T06:39M1.2 & S14E23 & 12403 & \cme\tablefootmark{c} & 07:12 (HALO; 06:18/06:35) & 547 & $+$ & \\
~ & SOL2015-08-22T13:17M2.2 & S15E19 & 12403 & C & $\times$ & $\times$ & $-$ & \\ 
~ & SOL2015-08-22T21:19M3.5 & S15E15 & 12403 & C\tablefootmark{c} & $\times$ & $\times$ & $-$ & \\ 
~ & \bf SOL2015-08-24T07:26M5.6 & S14E00 & 12403 & \cme\tablefootmark{b,c} & 08:48 (251; 07:30/07:45) & 272 & $-$ & ID 34 in Fig.~\ref{fig:dtw_major} \\
~ & SOL2015-08-24T17:40M1.0 & S15W04 & 12403 & C\tablefootmark{c} & $\times$ & $\times$ & $-$ & \\
\hline
~ & SOL2015-09-27T10:20M1.9 & S20W03 & 12422 & C\tablefootmark{c} & $\times$ & $\times$ & $-$ & \\
~ & SOL2015-09-27T20:54M1.0 & S21W16 & 12422 & C\tablefootmark{c} & $\times$ & $\times$ & $-$ & \\
~ & SOL2015-09-28T07:27M1.1 & S22W20 & 12422 & C\tablefootmark{c}  & $\times$ & $\times$ &  $-$ & \\ 
~ & SOL2015-09-28T13:01M1.1 & S20W16 & 12422 & C\tablefootmark{c} & $\times$ & $\times$ &  $-$ & \\ 
~ & \bf SOL2015-09-28T14:53M7.6 & S20W28 & 12422 & C\tablefootmark{b,c} & $\times$ & $\times$ & $-$ & ID 35 in Fig.~\ref{fig:dtw_major} \\
~ & SOL2015-09-29T03:41M1.1 & S20W36 & 12422 & C\tablefootmark{c} & $\times$ & $\times$ & $-$ & \\
~ & SOL2015-09-29T05:05M2.9 & S21W37 & 12422 & C\tablefootmark{c} & $\times$ & $\times$ & $-$ & \\ 
~ & SOL2015-09-29T05:53M1.0 & S20W30 & 12422 & C\tablefootmark{c} & $\times$ & $\times$ & $-$ & \\ 
~ & SOL2015-09-29T06:39M1.4 & S12W34 & 12422 & C\tablefootmark{c} & $\times$ & $\times$ & $-$ & \\
~ & SOL2015-09-29T11:09M1.6 & S21W27 & 12422 & C\tablefootmark{c} & $\times$ & $\times$ & $-$ & \\
\hline
~ & SOL2017-09-04T05:36M1.2 & S10W04 & 12673 & C, \gray\cme\tablefootmark{c} & 07:00 (\textcolor{gray}{356}; 05:21/06:00) & 188 & $-$ & \\ 
~ & SOL2017-09-04T15:11M1.5 & S10W08 & 12673 & C\tablefootmark{c} & $\times$ & $\times$ & $-$ & \\
~ & SOL2017-09-04T18:05M1.0 & S07W11 & 12673 & \cme\tablefootmark{c} & 19:00 (233; 18:35/18:11) & 597 & $+$ & \\
~ & SOL2017-09-04T18:46M1.7 & S09W11 & 12673 & \cme\tablefootmark{c} & 19:00 (233; 18:35/18:11) & 597 & $+$ & \\ 
~ & SOL2017-09-04T19:59M1.5 & S10W11 & 12673 & C\tablefootmark{c} & $\times$ & $\times$ & $-$ & \\ 
~ & \bf SOL2017-09-04T20:28M5.5 & S10W11 & 12673 & \cme\tablefootmark{c} & 20:36 (HALO; 20:21/20:14) & 418 & $+$ & ID 36 in Fig.~\ref{fig:dtw_major} \\
~ & SOL2017-09-04T22:10M2.1 & S09W12 & 12673 & \cme\tablefootmark{c} & $\times$ & $\times$ & $-$ & \\
~ & SOL2017-09-05T01:03M4.2 & S09W14 & 12673 & C\tablefootmark{c} & $\times$ & $\times$ & $-$ & \\
~ & SOL2017-09-05T03:42M1.0 & S09W15 & 12673 & C\tablefootmark{c} & $\times$ & $\times$ & $-$ & \\
~ & SOL2017-09-05T04:33M3.2 & S11W18 & 12673 & C\tablefootmark{c} & $\times$ & $\times$ & $-$ & \\
~ & SOL2017-09-05T17:37M2.3 & S10W23 & 12673 & \cme\tablefootmark{c} & 17:36 (216; 17:15/17:10) & 474 & $-$ & \\
~ & \bf SOL2017-09-06T08:57X2.2 & S08W32 & 12673 & C\tablefootmark{c} & $\times$ & $\times$ & $-$ & ID 37 in Fig.~\ref{fig:dtw_major} \\ 
~ & \bf SOL2017-09-06T11:53X9.3 & S09W34 & 12673 & \cme\tablefootmark{c} & 12:24 (HALO; 12:01/12:01) & 1571 & $+$ & \\
~ & SOL2017-09-06T15:51M2.5 & S08W36 & 12673 & \cme\tablefootmark{c} & $\times$ & $\times$ & $-$ & \\
~ & SOL2017-09-06T19:21M1.4 & S08W38 & 12673 & C\tablefootmark{c} & $\times$ & $\times$ & $-$ & \\
~ & SOL2017-09-06T23:33M1.2 & S08W40 & 12673 & C\tablefootmark{c} & $\times$ & $\times$ & $-$ & \\
~ & SOL2017-09-07T04:59M2.4 & S07W45\tablefootmark{\href{f:hfc}{\ref{f:hfc}}} & 12673 & C & $\times$ & $\times$ & $-$ & \\
~ & SOL2017-09-07T09:49M1.4 & S08W47\tablefootmark{\href{f:hfc}{\ref{f:hfc}}} & 12673 & C & $\times$ & $\times$ & $-$ & \\ 
~ & \bf SOL2017-09-07T10:11M7.3 & S07W46\tablefootmark{\href{f:hfc}{\ref{f:hfc}}} & 12673 & E & 10:24 (254; 09:44/09:52) & 470 & $+$ & \\
~ & \bf SOL2017-09-07T14:20X1.3 & S11W49\tablefootmark{\href{f:hfc}{\ref{f:hfc}}} & 12673 & E & 15:12 (254; 13:52/14:31) & 433 & $-$ & \\ 
\hline
\end{longtable}
\tablefoot{
Flare-AR associations have been established using the \tablefoottext{\href{f:ssle}{\ref{f:ssle}}}{SolarSoft Latest Events}and/or the \tablefoottext{\href{f:hfc}{\ref{f:hfc}}}{Hinode flare catalog (explicitly indicated).} Major flares (GOES class M5 or larger) are highlighted by a bold face flare identifier. For clarification of flare type (confined "C", or eruptive "E") the \tablefoottext{\href{f:lasco}{\ref{f:lasco}}}{SOHO/LASCO CME catalog} has been used, subject to the requirement that the position angle (PA) of a possibly associated CME has to be consistent with the respective flare position and that the linearly extrapolated CME onset time ($t_{0,\rm{linear}}$) is within a time window of $\pm$60~minutes around the GOES flare start time. In addition, based on the assumption of an expanding EUV wave/dimming (denoted by a "$+$" in the former last column) being indicative of a CME's occurrence, the \tablefoottext{{\href{f:sde}{\ref{f:sde}}}}{Solar Demon}EUV wave and dimming detections were scanned. For consistency checks, the listings of \tablefoottext{a}{\cite{2016SoPh..291.1761H},} \tablefoottext{b}{\cite{2018ApJ...853..105B},} \tablefoottext{c}{\cite{2020ApJ...900..128L}} were used. Partially contradictory indicators are highlighted in gray color. 
}
\end{appendix}

\end{document}